\documentstyle[tighten,aps,epsf,eqsecnum,amsfonts,prd]{revtex}
\input epsf
\begin{document}

\draft

\title{Quantum-mechanical model of the Kerr-Newman black hole}

\author{Jarmo M\"akel\"a\footnote{Electronic address: jarmo.makela@phys.jyu.fi},  Pasi Repo\footnote{Electronic address: pasi.repo@phys.jyu.fi},
Markus Luomajoki\footnote{Electronic address: markus.luomajoki@phys.jyu.fi} and Johanna Piilonen\footnote{Electronic address: johanna.piilonen@phys.jyu.fi}}

\address{Department of Physics, University of Jyv\"askyl\"a, P.O. Box 35, FIN-40351 Jyv\"askyl\"a, Finland}

\date{September 12, 2000}

\maketitle

\begin{abstract}

We consider a Hamiltonian quantum theory of stationary spacetimes containing a Kerr-Newman black hole. The physical phase space of such spacetimes is just six-dimensional, and it is spanned by the mass $M$, the electric charge $Q$ and angular momentum $J$ of the hole, together with the corresponding canonical momenta. In this six-dimensional phase space we perform a canonical transformation such that the resulting configuration variables describe the dynamical properties of Kerr-Newman black holes in a natural manner. The classical  Hamiltonian written in terms of these variables and their conjugate momenta is replaced by the corresponding self-adjoint Hamiltonian operator and an eigenvalue equation for the Arnowitt-Deser-Misner (ADM) mass of the hole, from the point of view of a distant observer at rest, is obtained. In a certain very restricted sense, this eigenvalue equation may be viewed as a sort of "Schr\"odinger equation of black holes". Our "Schr\"odinger equation" implies that the ADM mass, electric charge and angular momentum spectra of black holes are discrete, and the mass spectrum is bounded from below. Moreover, the spectrum of the quantity $M^2-Q^2-a^2$, where $a$ is the angular momentum per unit mass of the hole, is strictly positive when an appropriate self-adjoint extension is chosen. The WKB analysis yields the result that the large eigenvalues of $M$, $Q$ and $a$ are of the form $\sqrt{2n}$, where $n$ is an integer. It turns out that this result is closely related to Bekenstein's proposal on the discrete horizon area spectrum of black holes.

\end{abstract}

\pacs{Pacs number(s): 04.70.Dy, 04.20.Fy, 04.60.Ds, 04.60.Kz}


\section{Introduction}
\label{sec:Intro}

Black holes are among the simplest and most beautiful objects in the universe. They are made of spacetime and electromagnetic field only, and they have just three classical degrees of freedom which may be taken to be the mass $M$, electric charge $Q$ and the angular momentum $J$ of the hole.

Although the number of classical degrees of freedom of black holes is just three, however, one expects that there is an enormous number of quantum mechanical degrees of freedom associated with black holes. During some recent years, string theory and loop theoretic approaches to quantum gravity have greatly improved our understanding on the nature of these quantum mechanical degrees of freedom~\cite{Str,Ash}.

As it happens, there is a resemblance between black holes and hydrogen atoms. Like a black hole, a hydrogen atom has just three classical degrees of freedom. Indeed, the system looks very simple: An electron whirls around the proton, and the classical degrees of freedom may be taken to be the $x-$, $y-$ and $z-$coordinates of the electron. Quantum field theoretical investigations reveal, however, an enormous number of quantum mechanical degrees of freedom associated with virtual electron-positron pairs and photons. Still, the quantum mechanical properties of the hydrogen atom may be described, as an excellent approximation, by its non-relativistic Schr\"odinger equation which takes into account the three classical degrees of freedom only.

The resemblance between black holes and hydrogen atoms gives rise to an interesting question of a possibility to construct a quantum-mechanical model of a black hole which, although it takes into account the three classical degrees of freedon of black holes only, nevertheless describes their quantum mechanical properties with a reasonable accuracy. In this paper we shall consider one such model of black holes. Of course, even classical black hole spacetimes may perform all sorts of vibrations and oscillations which provide them with an enormous number of additional degrees of freedom, but in this paper we are interested in stationary black holes only. In other words, we are quantizing the stationary sector of black hole spacetimes, and such a sector is characterized by just three classical degrees of freedom~\cite{Heu}.

Our model is based on an observation that even stationary black hole spacetimes have {\it dynamics}. More precisely, even stationary black hole spacetimes have a region which does not admit a timelike Killing vector field. This  means that in a certain spacetime region the black hole spacetime geometry evolves in time no matter how we choose the time coordinate. It is this time evolution of black hole spacetime geometry on which we focus our attention and which, in our model, is responsible for the quantum mechanical properties of black holes.

To see what this means consider, as an example, the simplest possible black hole, the Schwarzschild black hole. It has the spacetime metric
\begin{equation}
ds^2=-\left( 1-\frac{2M}{r}\right) dt^2+\frac{dr^2}{ 1-\frac{2M}{r}}+r^2\left(d\theta ^2+\sin ^2 \theta d\phi ^2\right) \ .
\label{Schmetric}
\end{equation} 
One observes that when $r<2M$, the coordinate $r$ becomes timelike, and because spacetime geometry inside the event horizon depends on $r$, it evolves in time. In that region $r$ describes the radius of the wormhole throat of the black hole. In a more precise manner the fact that spacetime inside the event horizon really has dynamics in its geometry can be seen if one considers the conformal diagram of Kruskal spacetime: When $r<2M$ one cannot move in any timelike direction without changing $r$, and therefore the geometry of the spacelike hypersurfaces of spacetime. In Reissner-Nordstr\"om and Kerr-Newman black hole spacetimes the dynamical region lies in the intermediate region between the outer and the inner horizons of the hole.

In this paper we consider the Hamiltonian quantization of Kerr-Newman black hole spacetimes in such a manner that, in the classical level, the phase space coordinates of the theory describe the dynamics of the intermediate region between the horizons in a natural way. The Kerr-Newman solution is a specific solution to Einstein's and Maxwell's equations in vacuum. Because of that, we begin our investigations in Section~\ref{sec:Hamfor} by a general study of the Hamiltonian formulation of the Einstein-Maxwell theory, paying particular attention to the boundary terms which are needed in asymptotically flat electrovacuum spacetimes, such as Kerr-Newman spacetimes, to make the Hamiltonian formulation consistent. We shall see in later sections that these boundary terms play a most fundamental role in the quantum theory of Kerr-Newman black holes. In Section~\ref{sec:Bound} we calculate these boundary terms for maximally extended Kerr-Newman spacetimes. It turns out that from the boundary terms one can read off the mass, electric charge and angular momentum of the black hole.

The study of the classical Hamiltonian dynamics of Kerr-Newman black hole spacetimes is performed in Section~\ref{sec:Hamdyn}. Basically, our study is based on an important theorem proved by Regge and Teitelboim~\cite{Regge}. This theorem states, essentially, that the physical Hamiltonian of asymptotically flat spacetime with matter fields can be gained if we first solve the classical constraints, and then substitute the solutions to the constraints, in terms of the physical phase space coordinates of the theory, to the boundary terms at asymptotic spacelike infinity. At the first stage we take the phase space coordinates to be the mass $M$, the electric charge $Q$ and the angular momentum $J$ of the hole, together with the corresponding canonical momenta $p_M,\ p_Q$ and $p_J$, and we write the sum of the boundary terms, and hence the classical Hamiltonian, in terms of these phase space coordinates. It is unclear whether the assumptions of the Regge's and Teitelboim's theorem are valid for Kerr-Newman spacetime and the variables $M,\ Q,\ J,\ p_M,\ p_Q$ and $p_J$, but we accept this as an unproved hypothesis and see where it takes us. At the second stage we perform a canonical transformation from the variables $M,\ Q$ and $J$ and their canonical momenta to the new variables and their canonically conjugate momenta which describe better the dynamics of the intermediate region between the horizons of the Kerr-Newman black hole. In terms of these phase space variables we write the classical Hamiltonian of Kerr-Newman spacetimes in a specific foliation where the flat Minkowski time coordinate of an asymptotic observer at rest at a faraway infinity coincides with the the proper time of a freely falling observer at the throat of the black hole. An explicit example of such a foliation is presented in Appendix~\ref{app:B}.

In Section~\ref{sec:Quant} we proceed to quantization. A straightforward replacement of the classical Hamiltonian by the corresponding self-adjoint Hamiltonian operator yields an equation which, in a certain very restricted sense, may be considered as a sort of "Schr\"odinger equation of black holes". That equation is the main result of this paper. In the natural units, where $\hbar = c= G=4\pi\epsilon _0 =1$, and when a particular operator ordering has been chosen, it can be written, in terms of the configuration variables $R,\ u$ and $v$ of the theory, as:
\begin{equation}
\frac{1}{2R}\left( -\frac{\partial ^2}{\partial R^2} -\frac{\partial ^2}{\partial u^2} -\frac{\partial ^2}{\partial v^2} +R^2+u^2+v^2\right)\Psi = M\Psi \ ,
\label{Result}
\end{equation}
where $\Psi = \Psi (R,u,v)$ is the wave function of the Kerr-Newman black hole.

Besides of being an equation of considerable simplicity,  our "Schr\"odinger equation" has many interesting consequences. For instance, it predicts that the mass, electric charge and angular momentum spectra of black holes are discrete. In particular, the mass spectrum is bounded from below and can be made positive by means of an appropriate choice of a self-adjoint extension. As a matter of fact, one may prove even more than that: It is possible to choose operator orderings and self-adjoint extensions such that the spectrum of the quantity
$$
M^2-Q^2-a^2 \ ,
$$ 
where $a:=\frac{J}{M}$ is the angular momentum per unit mass, is {\it strictly positive}. Regarding Hawking radiation, this is very interesting result: It implies that a non-extreme black hole can never become, by means of Hawking radiation, an extreme hole. This result is in agreement with the third law of black hole thermodynamics, and is therefore a strong argument in favor of the physical validity of our model.

At the high end of the spectrum, we find that the eigenvalues of the sum of the areas of the horizons of the hole which we shall call, for the sake of convenience, the {\it total area} of the black hole, are of the form
\begin{equation}
A^{\rm tot}_n =n\cdot 16\pi l_{\rm Pl}^2 \ ,
\label{areaspectrum}
\end{equation} 
where $n=1,2,3,\dots$ and $l_{\rm Pl}:= \sqrt{\frac{\hbar G}{c^3}}$ is the Planck length. As such our result is closely related, although not quite identical to Bekenstein's proposal~\cite{Bek1,Muk,Bek2,Bek3,Maz,Kog,Mag,Lousto,Pel,Louko,Bar,Kas1,Mak1,Hod1,Mak2,Hod2,Hod3,Kas2,Bojo,Ah,Gar}. According to that proposal, the spectrum of the outer horizon of the hole is of the form
\begin{equation}
A_n =n\cdot \gamma l_{\rm Pl}^2 \ ,
\label{Bproposal}
\end{equation} 
where $n=1,2,3,\dots$ and $\gamma$ is pure number of order one. Arguments in favor of the claim that it is perhaps not the area of the exterior horizon but the sum of the areas which should have an equal spacing in its spectrum will be given in Section~\ref{sec:Conc}.

Finally, our model implies a very interesting discrete spectrum for the angular momentum of Kerr-Newman black holes. For uncharged holes near extremality, the angular momentum eigenvalues are of the form
\begin{equation}
J_m =m\hbar \ ,
\label{J}
\end{equation} 
where $m=0,\pm 2,\pm 4,\dots$.

\section{Hamiltonian formulation of Einstein-Maxwell theory}
\label{sec:Hamfor}

The Einstein-Maxwell theory is a theory of electromagnetic field interacting with gravitational field. In this section we shall develope the Hamiltonian formulation of such a theory in all details, paying particular attention to the boundary terms appearing in asymptotically flat spacetimes as a consequence of the requirement of internal consistency of the theory.

The action of the Einstein-Maxwell theory can be written, in general, as
\begin{equation}
S=\frac{1}{16 \pi}\int d^4x \sqrt{-g}\left(\!^{(4)}\!R - F_{\mu \nu }F^{\mu \nu }\right) + ({\mathrm boundary \ terms})\ .
\label{EMaction}
\end{equation}
In this equation the integration is performed over the whole four-dimensional spacetime. $g$ is the determinant of the spacetime metric $g_{\mu \nu }$, and
\begin{equation}
F_{\mu \nu }:=\partial _{\mu}A_{\nu}- \partial _{\nu}A_{\mu}
\end{equation}
is the electromagnetic field tensor. $A_{\mu}$ is the electromagnetic vector potential. $\!^{(4)}\!R$ is the four-dimensional scalar curvature.

As it is well known, we can write the action~(\ref{EMaction}) as
\begin{equation}
S=S_{\Sigma}^{\mathrm grav}+S_{\Sigma}^{\mathrm em}+S_{\partial \Sigma}^{\mathrm grav}+S_{\partial \Sigma}^{\mathrm em}\ ,
\label{totact}
\end{equation}
where
\begin{mathletters}
\label{actions}
\begin{eqnarray}
S_{\Sigma}^{\mathrm grav}&:=&-\frac{1}{16\pi }\int dt \int _{\Sigma} d^3x \sqrt{q}N(K_{ab}K^{ab}-K^2+{\mathcal R}) \ ,
\label{gravact} \\
S_{\Sigma}^{\mathrm em}&:=&\frac{1}{16\pi }\int dt \int _{\Sigma} d^3x \sqrt{q}NF_{\mu \nu }F^{\mu \nu }\  ,
\label{emact}
\end{eqnarray}
\end{mathletters}
and $S_{\partial \Sigma}^{\mathrm grav}$ and $S_{\partial \Sigma}^{\mathrm em}$ are boundary terms associated with spacelike asymptotic infinities of asymptotically flat spacetimes. In Eqs.~(\ref{actions}) the spatial integration is performed over the whole spacelike hypersurface $\Sigma$ of spacetime where the time $t$ is constant. $K_{ab}$ is the exterior curvature tensor on that hypersurface, $K$ its trace, and ${\mathcal R}$ is the three-dimensional scalar curvature on that hypersurface. $N$ is the lapse function and $q$ is the determinant of the metric $q_{ab}$ on the hypersurface $\Sigma$.

The properties of the action $S_{\Sigma}^{\mathrm grav}$ are well known. Consider now the action  $S_{\Sigma}^{\mathrm em}$ of Eq.~(\ref{emact}). To begin with, consider first the case where the spacetime metric can be written as
\begin{equation}
ds^2 = - dt^2 + q_{ab}dx^adx^b \ .
\end{equation}
In other words, we have chosen time orthogonal coordinates, where the lapse $N \equiv 1$, and the shift $N^a$ vanishes identically. In these coordinates we can write
\begin{equation}
S_{\Sigma}^{\mathrm em}=\int dt \int _{\Sigma} d^3x{\mathcal L}^{\mathrm em}\ ,
\label{emact'}
\end{equation}
where
\begin{equation}
{\mathcal L}^{\mathrm em}:=\frac{1}{16\pi}\sqrt{q}\left\lbrace2q^{ab}\left[\dot A_a \dot A_b -2\dot A_a(\partial _bA_0) + (\partial _a A_0)(\partial _b A_0)\right]- \ ^{(3)}\!F_{ab}\! ^{(3)}\!F^{ab}\right\rbrace
\label{emlag}
\end{equation}
is the electromagnetic Lagrangian in curved spacetime. The dot means time derivative and we have defined
\begin{mathletters}
\label{emfields}
\begin{eqnarray}
\ ^{(3)}\!F_{ab}&:=&\partial _a A_b -\partial _b A_a \ ,
\label{emfield} \\
\ ^{(3)}\!F^{ab}&:=&q^{am}q^{bn}\,^{(3)}\!F_{mn} \ .
\label{emfieldup}
\end{eqnarray}
\end{mathletters}
The canonical momentum conjugate to $A_a$ is
\begin{equation}
p^a:=\frac{\partial {\mathcal L}^{\mathrm em}}{\partial \dot A_a} = \frac{\sqrt{q}}{4\pi}q^{as}\left(\dot A_s-\partial_sA_0\right) = \frac{\sqrt{q}}{4\pi}q^{as}F_{0s} \ .
\label{pa}
\end{equation}
This relation can be inverted, and we have
\begin{equation}
\dot A_b = \frac{4\pi}{\sqrt{q}}p_b+\partial_b A_0 \ , 
\label{invpa}
\end{equation}
where we have defined
\begin{equation}
p_b := q_{ab}p^a\ . 
\label{pb}
\end{equation}
In terms of $p^a$ we can write the electromagnetic Lagrangian as
\begin{equation}
{\mathcal L}^{\mathrm em} = p^a\dot A_a -\left[ \frac{2\pi}{\sqrt{q}}q_{ab} p^ap^b+p^a (\partial_a A_0) + \frac{\sqrt{q}}{16\pi}  \!^{(3)}\!F_{ab}\! ^{(3)}\!F^{ab}  \right] \ .
\label{Lem}
\end{equation}
Hence, we get
\begin{equation}
S_{\Sigma}^{\mathrm em}=\int dt \int _{\Sigma} d^3x\left[ p^a\dot A_a-{\mathcal H}^{\mathrm em}+ A_0(\partial_ap^a)  \right]\ ,
\label{emact''}
\end{equation}
where
\begin{equation}
{\mathcal H}^{\mathrm em} := \frac{2\pi}{\sqrt{q}}q_{ab}p^ap^b+\frac{\sqrt{q}}{16\pi} \!^{(3)}\!F_{ab}\! ^{(3)}\!F^{ab}\ .
\label{Hamem}
\end{equation}
In Eq.~(\ref{emact''}) we have dropped the term $\frac{1}{2}\int dt \int _{\Sigma} d^3x \partial_a(A_0p^a)$ which can be inverted to a boundary term.

We now include the lapse and the shift to our formulation. To include the lapse we replace $dt$ by
\begin{equation}
dt'=Ndt \ ,
\label{lapse}
\end{equation}
and, because $A_0$ transforms to
\begin{equation}
A'_0\ = \frac{\partial x^\mu}{\partial x'^0}A_\mu \ ,
\label{transofA_0}
\end{equation}
we find that for general lapse but vanishing shift the electromagnetic action is
\begin{equation}
S_{\Sigma}^{\mathrm em}=\int dt \int _{\Sigma} d^3x\left[ p^a\dot A_a-N{\mathcal H}^{\mathrm em}+ A_0(\partial_ap^a)  \right]\ .
\label{emact'''}
\end{equation}

Inclusion of a non-vanishing shift is a bit more tricky. We replace $dx^a$ by
\begin{equation}
dx'^a=dx^a+N^a dt \ ,
\label{shiftvector}
\end{equation}
from which it follows that $A_0$ is replaced by
\begin{equation}
A'_0=A_0 - N^s A_s \ .
\label{A'0}
\end{equation}
Moreover, at the hypersurface where $x^0=t+dt$, $A_a$ is replaced by
\begin{equation}
A'_a = \frac{\partial x^s}{\partial x'^a}A_s (t+dt, x^b-N^bdt) = A_a+\dot A_a dt -(\partial_s A_a)N^s dt-(\partial_a N^s)A_sdt \ .
\label{A'0'}
\end{equation}
Hence, we find that $\dot A_a$ must be replaced by 
\begin{equation}
\dot A'_a = \dot A_a  -(\partial_s A_a)N^s -(\partial_a N^s)A_s \ .
\label{dotA'0}
\end{equation}
Substituting Eqs.~(\ref{A'0}) and~(\ref{dotA'0}) into Eq.~(\ref{emact'''}) we obtain an expression for the electromagnetic action in the presence of non-vanishing shift:
\begin{equation}
S_{\Sigma}^{\mathrm em}=\int dt \int _{\Sigma} d^3x\left[ p^a\dot A_a-N{\mathcal H}^{\mathrm em}-N^s{\mathcal H}_s^{\mathrm em}+ A_0(\partial_ap^a)  \right]\ ,
\label{emaction}
\end{equation}
where we have defined
\begin{equation}
{\mathcal H}_s^{\mathrm em} := p^a\ ^{(3)}\!F_{sa}\ ,
\label{amdiffco}
\end{equation}
and we have ignored the term $\int dt\int _{\Sigma} d^3 x\partial _a(A_sN^sp^a)$.

We are now prepared to write down the whole Einstein-Maxwell action without boundary terms. The gravitational part $S_\Sigma^{\mathrm grav}$ is a mere ADM action
\begin{equation}
S_\Sigma^{\mathrm grav} = \int dt\int _\Sigma d^3 x \left( p^{ab}\dot q_{ab} -N {\mathcal H}^{\mathrm grav} -N^s{\mathcal H}^{\mathrm grav}_s\right) \ ,
\label{gravaction}
\end{equation}
where
\begin{mathletters}
\label{gravconstraints}
\begin{eqnarray}
{\mathcal H}^{\mathrm grav}&:=& \frac{1}{2}(16\pi)G_{abcd}p^{ab}p^{cd}+\frac{1}{16\pi}\sqrt{q}{\mathcal R} \ , \\
\label{gravHamco}
{\mathcal H}^{\mathrm grav}_s &:=& -2p_{s\ \vert a}^{\ a}
\label{gravdiffco} \ ,
\end{eqnarray}
\end{mathletters}
and 
\begin{equation}
p^{ab} := -\frac{1}{16\pi}\sqrt{q}\left(K^{ab}-q^{ab}K\right)
\label{momentumpab}
\end{equation}
is the canonical momentum conjugate to $q_{ab}$.
\begin{equation}
G_{abcd} := -\frac{1}{\sqrt{q}}\left(q_{ab}q_{cd}-q_{ac}q_{bd}-q_{ad}q_{bc}\right)
\label{Wheelermetric}
\end{equation}
is the Wheeler-De Witt metric. Putting the actions~(\ref{emaction}) and ~(\ref{gravaction}) together we get the Einstein-Maxwell action
\begin{equation}
S_\Sigma^{\mathrm grav} = \int dt\int _\Sigma d^3 x \left( p^{ab}\dot q_{ab}+p^a\dot A_a -N {\mathcal H} -N^s{\mathcal H}_s-A_0{\mathcal G}\right) \ ,
\label{Action}
\end{equation}
where 
\begin{equation}
{\mathcal H}:= {\mathcal H}^{\mathrm grav} + {\mathcal H}^{\mathrm em}
\label{Hamcosum}
\end{equation}
is the Hamiltonian constraint,
\begin{equation}
{\mathcal H}_s:= {\mathcal H}^{\mathrm grav}_s + {\mathcal H}^{\mathrm em}_s
\label{Diffcosum}
\end{equation}
is the diffeomorphism constraint, and 
\begin{equation}
{\mathcal G}:= -\partial_a p^a
\label{Gaussco}
\end{equation}
is the Gaussian constraint.

We shall now consider asymptotically flat spacetimes. In those kind of spacetimes we must include certain boundary terms, since we cannot assume the variations of the dynamical variables and their canonical momenta to vanish at asymptotic infinity. In what follows, we shall take the asymptotic coordinates at spatial infinity to be Cartesian coordinates.

First of all, of course, we have the ADM boundary term~\cite{Regge,Kuc}
\begin{equation}
S^{\mathrm ADM}_{\partial \Sigma} = -\int dt N^+(t) E_{\mathrm ADM}(t) \ ,
\label{ADMboundary}
\end{equation}
where 
\begin{equation}
N^+(t) := \lim_{r\rightarrow \infty} N(t,x^a)
\label{lapse+}
\end{equation}
is the lapse function at the asymptotic spatial infinity, and 
\begin{equation}
E_{\mathrm ADM} := \lim_{r \rightarrow \infty} \frac{1}{16\pi}\oint \left( \frac{\partial h_{mn}}{\partial x^n} - \frac{\partial h_{nn}}{\partial x^m}\right) dS^m
\label{Eadm}
\end{equation}
is the ADM energy of spacetime. In Eq.~(\ref{Eadm}) $h_{mn}$ denotes a spatial component of the linearized gravitational field in asymptotic Cartesian coordinates. More precisely, we have assumed spatial coordinates to become Cartesian in asymptotic spacelike infinity, and in these coordinates we have written the spacetime metric as $g_{\mu \nu}=\eta _{\mu \nu}+h_{\mu \nu}$, where $\eta _{\mu \nu}={\mathrm diag}(-1, 1, 1, 1)$ is the flat spacetime metric.

In addition to the ADM boundary term, which is a term associated with time evolution at asymptotic infinity, we have, for the non-vanishing shift at spatial infinity, boundary terms associated with asymptotic spatial translations. Variation of the action~(\ref{Action}) with respect to the momentum $p^{ab}$ conjugate to $q_{ab}$ brings along a term~\cite{Regge}
$$
2\int dt \int_\Sigma d^3 x\left(N_a\delta p^{ab}\right)_{\vert b} \ ,
$$
which must be cancelled at infinity. Hence we need a boundary term
\begin{equation}
S^{\mathrm trans}_{\partial \Sigma} := - \int dt N^+_a(t)P^a_{\mathrm ADM}(t) \ ,
\label{transb}	
\end{equation}
where
\begin{equation}
N^+_a(t) := \lim_{r\rightarrow \infty} N_a(t,x^a)
\label{shift'}
\end{equation}
is the shift at infinity, and
\begin{equation}
P^a_{\mathrm ADM} := -\lim_{r\rightarrow \infty} 2\oint p^{ab}dS_b
\label{ADMmomentum}
\end{equation}
is the ADM momentum of spacetime.

So far we have considered terms related to pure gravity. We still have to include boundary terms related to electromagnetism. First of all, we observe that variation of the action with respect to the momentum $p^a$ conjugate to $A_a$ brings along a term
$$
\int dt \int_\Sigma d^3 x\partial_a \left(A_0\delta p^a\right) \ ,
$$
which must be cancelled at infinity. Hence, we need an electromagnetic boundary term
\begin{equation}
S^{\mathrm em}_{\partial \Sigma} := -\int dt A^+_0(t)Q(t) \ ,
\label{emb}
\end{equation}
where
\begin{equation}
A^+_0(t) := \lim_{r\rightarrow \infty} A_0(t,x^a)
\label{A0+}
\end{equation}
is the electric potential at infinity, and
\begin{equation}
Q := -\lim_{r\rightarrow \infty} \oint p_adS^a
\label{Q}
\end{equation}
is the electric charge of spacetime.

We are now prepared to write the whole Einstein- Maxwell action, with appropriate boundary terms. We get
\begin{equation}
S_{\Sigma} = \int dt\int _\Sigma d^3 x \left( p^{ab}\dot q_{ab}+p^a\dot A_a -N {\mathcal H} -N^s{\mathcal H}_s-A_0{\mathcal G}\right) -\int dt  \left( N^+(t) E_{\mathrm ADM}(t) + N^+_a(t)P^a_{\mathrm ADM}(t) +A^+_0(t)Q(t) \right) \ .
\label{Totaction}
\end{equation}
Because of that, the total Hamiltonian of the Einstein-Maxwell theory is
\begin{equation}
H = \int _\Sigma d^3 x \left(  N {\mathcal H} +N^s{\mathcal H}_s+A_0{\mathcal G}\right) + N^+(t) E_{\mathrm ADM}(t) + N^+_a(t)P^a_{\mathrm ADM}(t) +A^+_0(t)Q(t)  \ .
\label{TotHamiltonian}
\end{equation}
Hence, one is left with the last three terms only when the classical constrainsts
\begin{mathletters}
\label{constraints}
\begin{eqnarray}
{\mathcal H}&=& 0 \ , \\
\label{Hamco}
{\mathcal H}_s &=& 0 \ , \\
\label{diffco}  
{\mathcal G}&=& 0 
\label{Gaussco'}
\end{eqnarray}
\end{mathletters}
are satisfied.

\section{Boundary terms in Kerr-Newman spacetime}
\label{sec:Bound}

As we saw in the previous section, one must include, in asymptotically flat spacetimes, certain boundary terms in order to make the variational principle consistent. Of particular interest are the boundary terms in Kerr-Newman spacetime, the most general black hole spacetime. In this section we shall calculate these boundary terms.

To begin with, we write down the Kerr-Newman line element in Boyer-Lindquist coordinates~\cite{Wald}:
\begin{equation}
ds^2 = -  \frac{\Delta -a^2\sin ^2\theta}{\Sigma} dt^2 -
\frac{2a\sin^2\theta(r^2+a^2-\Delta)}{\Sigma}dtd\phi+
\frac{(r^2+a^2)^2-\Delta a^2\sin^2\theta}{\Sigma}\sin^2\theta d\phi^2+
\frac{\Sigma}{\Delta}dr^2+
\Sigma d\theta^2 \ ,
\label{KNelement}
\end{equation}
where
\begin{mathletters}
\begin{eqnarray}
\Sigma &:=& r^2+a^2\cos^2\theta \ ,\\
\Delta &:=& r^2+a^2+Q^2-2Mr \ .
\label{sig&de}
\end{eqnarray}
\end{mathletters}
In these equations, $M$ is the ADM mass of the hole, $Q$ its charge, and $a$ is the angular momentum per unit mass. To calculate the boundary terms we must approximate the line element~(\ref{KNelement}) at asymptotic infinity, where $r\rightarrow \infty$, when only leading order terms are taken into account:
\begin{equation}
ds^2 \approx -\left(1-\frac{2M}{r}\right)dt^2-\frac{4J\sin^2\theta}{r}dtd\phi+r^2\sin^2\theta d\phi^2+\left(1+\frac{2M}{r}\right)dr^2+r^2d\theta^2 \ ,
\label{appKNelement}
\end{equation}
where $J:=Ma$ is the angular momentum of the hole. In Cartesian coordinates this expression takes the form
\begin{equation}
ds^2 \approx -\left(1-\frac{2M}{r}\right)dt^2-\frac{4J}{r^3}(xdy-ydx)dt+\left(1+\frac{2M}{r}\right)(dx^2+dy^2+dz^2) \ ;
\label{appKNelement'}
\end{equation}
here $r$ is not the same $r$ as in Eq.~(\ref{appKNelement}): In Eq.~(\ref{appKNelement}) $r$ is one of the Boyer-Lindquist coordinates, whereas in Eq.~(\ref{appKNelement'}) $r$ is defined to be equal to $(x^2+y^2+z^2)^{1/2}$.

We now proceed to evaluate the boundary terms. When evaluating the boundary terms the first task is to fix the coordinate system far away from the black hole. In other words, we must fix the lapse $N$ and the shift $N^a$. In this paper we choose a faraway coordinate system which revolves, with respect to the Cartesian coordinates $x$, $y$ and $z$, with an extremely small angular velocity $\omega$ around the $z$-axis. (We must choose $\omega$ to be extremely small since otherwise the velocities of faraway observers would exceed the velocity of light.) Because, in flat space, the velocity of an observer at the point $\vec r = x\hat i+y\hat j+z\hat k$ revolving with angular velocity $\vec \omega$ is
\begin{equation}
\vec v = \vec \omega \times \vec r\ ,
\label{velocity}
\end{equation}
and because, in Cartesian coordinates, $N^a$ represents the $a$-component of velocity, we find that 
\begin{equation}
N^a = \varepsilon^a_{bc} \omega^bx^c \ ,
\label{shift}
\end{equation}
where $\varepsilon^a_{bc}$ is the Levi-Civita symbol such that $\varepsilon_{123}=1$.

What sort of boundary terms do show up with this kind of a choice of the shift? To begin with, we recall from Section~\ref{sec:Hamfor} that variation of the momentum $p^{ab}$ conjugate to $q_{ab}$ brings along a term
$$
2\int dt\int d^3x(N_a \delta p^{ab})_{\vert b} \ ,
$$
which must be cancelled at infinity. If the shift $N^a$ is chosen as in Eq.~(\ref{shift}), we must therefore bring along a boundary term
\begin{equation}
S^{\mathrm rev}_{\partial \Sigma} = -2\varepsilon_{abc}\int dt\omega^b\lim_{r\rightarrow \infty} \oint x^cp^{an}dS_n \ ,
\label{boundrot}
\end{equation}
which replaces the boundary term $S^{\mathrm trans}_{\partial \Sigma}$ of Eq.~(\ref{transb}).

Now, when calculating the boundary term $S^{\mathrm rev}_{\partial \Sigma}$ of Eq.~(\ref{boundrot}) we should, of course, first perform a coordinate transformation where the spacetime metric~(\ref{appKNelement'}) is replaced by the corresponding expression in revolving coordinates, and then proceed to calculate the boundary term by using this expression. However, when the faraway coordinate system revolves very slowly, we are interested in terms linear in $\omega$ only. Taking into account the transformation in the expression of the metric would produce terms quadratic in $\omega$, which can be neglected. Hence, we are allowed to calculate the boundary term~(\ref{boundrot}) by using the metric~(\ref{appKNelement'}). This calculation has been performed in details in Appendix~\ref{app:A}, and we get, because $\vec \omega = \omega \hat k$:
\begin{equation}
S^{\mathrm rev}_{\partial \Sigma} = -\int dt\, \omega J \ .
\label{boundrot'}
\end{equation}

We must still calculate the ADM boundary term~(\ref{ADMboundary}) as well as the electromagnetic boundary term~(\ref{emb}). The ADM boundary term of Kerr-Newman spacetime is, for arbitrary lapse $N^+$ at infinity,
\begin{equation}
S^{\mathrm ADM}_{\partial \Sigma} = - \int dt N^+M \ .
\label{boundADM}
\end{equation}
To calculate the electromagnetic boundary term we first recall that for Kerr-Newman solution the only non-zero components of $A_\mu$ in Boyer-Lindquist coordinates are
\begin{mathletters}
\label{Amyy}
\begin{eqnarray}
A_t &=& - \frac{Qr}{\Sigma} \ , \\
A_\phi &=& \frac{Qar}{\Sigma}\sin^2\theta \ .
\end{eqnarray}
\end{mathletters}
Using Eqs.~(\ref{pa}),~(\ref{A'0}) and~(\ref{dotA'0}) one finds that for general lapse and shift one can write $p^a$, the canonical momentum conjugate to $A_a$ as
\begin{equation}
p^a = \frac{1}{N}\frac{\sqrt{q}}{4\pi}q^{as}\left( F_{0s}-N^{b(3)}F_{bs}\right)\  .
\label{momentumtoA}
\end{equation}
This expression, together with Eqs.~(\ref{Amyy}), implies that in Boyer-Lindquist coordinates the only surviving component of $p^a$ is $p^r$ which, in the leading order, can be written very far away from the hole as
\begin{equation}
p^r = -\frac{Q}{4\pi r^2}+{\mathcal O} \left( r^{-3}\right)\  .
\label{pr}
\end{equation}
Hence, the electromagnetic boundary term~(\ref{emb}) is
\begin{equation}
S^{\mathrm em}_{\partial \Sigma} = - \int dt A_0^+Q\ ,
\label{boundem}
\end{equation}
as expected. The slow rotation of the asymptotic coordinate system will change the ADM and the electromagnetic boundary terms a bit but the resulting corrections will be of the order of ${\mathcal O} (\omega ^2)$ and can therefore be neglected.

\section{Hamiltonian dynamics of Kerr-Newman spacetimes}
\label{sec:Hamdyn}

\begin{figure}
    \begin{center}
      \leavevmode
      \epsfysize=90mm
      \epsfbox{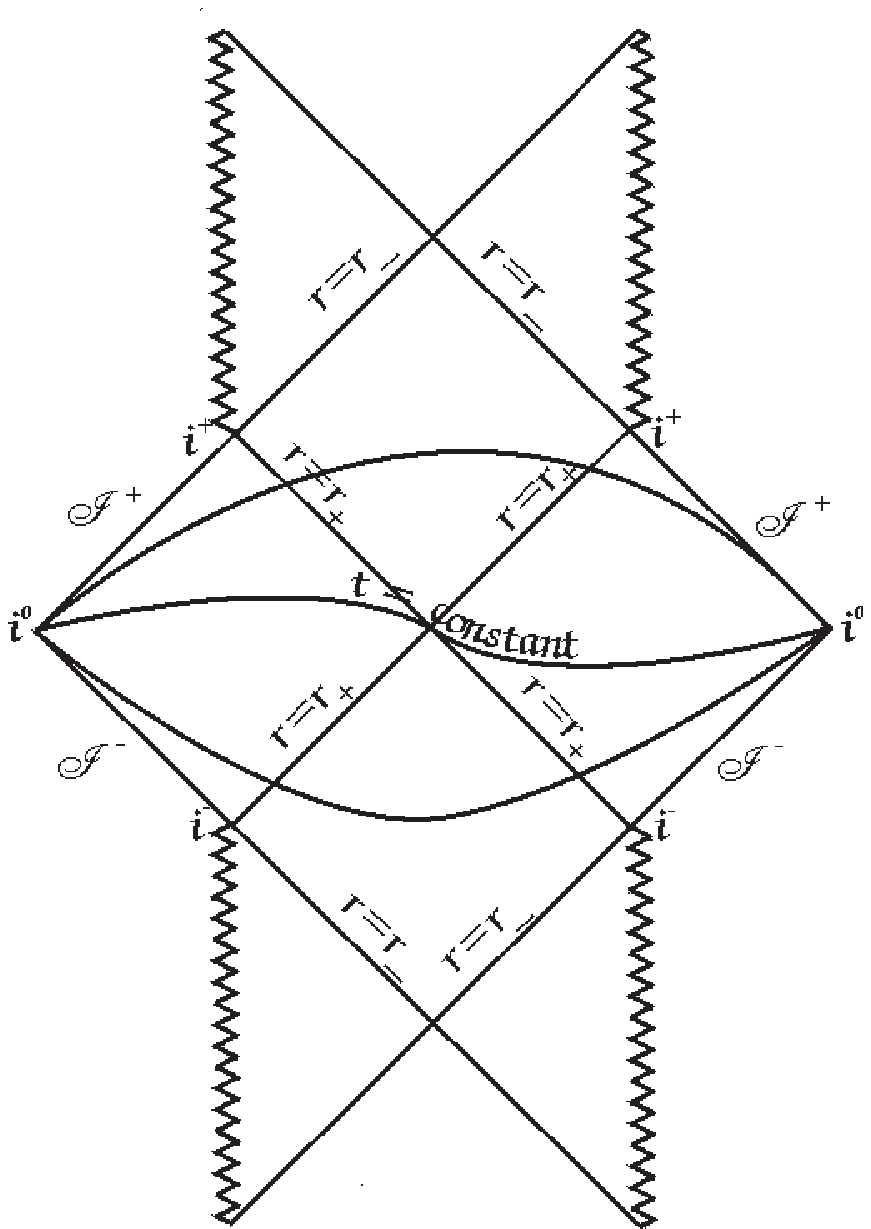}
    \end{center}
  \caption{The conformal diagram of Kerr-Newman spacetimes. Our spacelike hypersurfaces $t=constant$ begin their life at the past $r=r_-$ hypersurface, then go through the bifurcation point, and finally end their life at the future $r=r_-$ hypersurface.}
  \label{fig1}
 \end{figure}

We shall now proceed to the study of the Hamiltonian dynamics of maximally extended Kerr-Newman spacetimes. To begin with, consider a foliation of such spacetimes into space and time. Obviously, we want the spacelike hypersurfaces where the time $t =constant$ to cover as great a portion of spacetime as possible. Maximally extended Kerr-Newman spacetimes have a periodic geometrical structure, and we pick up one such period~\cite{Wald}. We choose the spacelike hypersurfaces $t=constant$ such that they begin from the left hand side asymptotic infinity, then go through the interior regions of the hole in arbitrary ways, and finally end at the right hand side asymptotic infinity in the conformal diagram. However, such spacelike hypersurfaces cannot be pushed beyond the interior horizons where the Boyer-Lindquist coordinate
\begin{equation}
r=r_- := M-\sqrt{M^2-Q^2-a^2 \ },
\label{r-}
\end{equation}
since otherwise our hypersurfaces would fail to be spacelike. Hence our study of the Hamiltonian dynamics of Kerr-Newman spacetimes must be restricted to include, in addition to the left and the right exterior regions of the Kerr-Newman black hole, only such an interior region of the hole that lies between two successive $r=r_-$ hypersurfaces in the conformal diagram. Our spacelike hypersurface $t=constant$ therefore begins its life at the past $r=r_-$ hypersurface, then goes through the bifurcation point where
\begin{equation}
r=r_+ :=M+\sqrt{M^2-Q^2-a^2} \ ,
\label{r+}
\end{equation}
and finally ends its life at the future $r=r_-$ hypersurface (See Fig.~\ref{fig1}). Bearing this restriction in mind, we shall now go into the Hamiltonian dynamics of Kerr-Newman spacetimes.

The first task is to write the action with appropriate boundary terms. The problem is now that we have {\it two} asymptotic infinities, and at both of these infinities we have certain boundary terms. When this fact is taken into account, we find that the action takes the form
\begin{equation}
S=\int dt \int _{\Sigma}d^3x\left(p^{ab}\dot q_{ab} + p^a\dot A_a - N{\mathcal H}-N^s{\mathcal H}_s -A_0{\mathcal G}\right)
-\int dt\left[(N^+ + N^-)M + (A_0^+-A_0^-)Q+(\omega ^+ -\omega^-)J\right]\ .
\label{action}
\end{equation}
In this equation, quantities equipped with plus and minus, respectively, are quantities written at the right and the left asymptotic infinities. In particular, $\omega ^+$ and $\omega ^-$ are angular velocities of faraway coordinate systems around $z$-axis. Hence, the total Hamiltonian of Kerr-Newman spacetime is
\begin{equation}
H_{tot}=\int _{\Sigma}d^3x\left(N{\mathcal H}+N^s{\mathcal H}_s +A_0{\mathcal G}\right)
+(N^+ + N^-)M+(A_0^+-A_0^-)Q+(\omega ^+ - \omega ^-)J \ .
\label{totHam}
\end{equation}

Now, the problem with our Hamiltonian is that it contains an enormous number of independent degrees of freedom. Indeed, our Hamiltonian may be considered as a function of the hypersurface metric $q_{ab}$ at each point $x$ on the spacelike hypersurface $\Sigma$, together with the corresponding canonical momenta $p^{ab}$. However, the ultimate object of interest in this paper is canonical quantization of the stationary black hole sector of Einstein-Maxwell theory. Stationary black holes, in turn, are characterized by just three independent, classical degrees of freedom, and hence an enormous number of degrees of freedom must be truncated.

For non-rotating black hole spacetimes, the truncation process may be performed in the following manner: One first writes the action for asymptotically flat, spherically symmetric Einstein-Maxwell theory. After the Hamiltonian, diffeomorphism and Gaussian constraints have been solved, one is left with just four canonical degrees of freedom which can be taken to be the mass $M$ and the electric charge $Q$ of the hole, together with the corresponding canonical momenta $p_M$ and $p_Q$~\cite{Ah,Kuc,Wint}. A similar truncation could also be performed for rotating black holes: One begins with asymptotically flat Einstein-Maxwell theory with appropriate symmetries, solves the classical constraints, and is finally left with just six physical, canonical degrees of freedom which may be taken to be the mass $M$, the electric charge $Q$, and the angular momentum $J$ of the Kerr-Newman black hole, together with the corresponding canonical momenta $p_M$, $p_Q$ and $p_J$.

An important feature of the process explained above, in which the phase space becomes reduced in such a way that only the physical degrees of freedom are left, is that the resulting Hamiltonian, the so called reduced Hamiltonian, involves the boundary terms only. In particular, the reduced Hamiltonian of Kerr-Newmann spacetimes is now
\begin{equation}
H^{\mathrm red}=(N^+ + N^-)M + (A_0^+ -A_0^-)Q+(\omega ^+ - \omega ^- )J \ .
\label{redHam}
\end{equation}

As a matter of fact, the reduced Hamiltonian may be used as the real, physical Hamiltonian of the system. This was proved by Regge and Teitelboim~\cite{Regge}. They found that if one solves the classical constraints and then substitutes the solutions to the reduced Hamiltonian, then the correct equations of motion for the canonical variables are obtained. More precisely, they showed the following: One assumes that the variables $q_{ab}$ and $p^{ab}$ can be separated by a one to one, time independent, functionally differentiable canonical transformation in  two sets $(\varphi ^{\alpha}, \pi _{\alpha})$ and $(\psi ^A, \pi _A)$ in such a way that: 
\begin{itemize}
\item[a)]
The reduced Hamiltonian depends only on $\varphi ^{\alpha}$ and the $\pi _{\alpha}$ \ .

\item[b)] 
When the $\pi _{\alpha}$ are prescribed as functions $p_{\alpha}$ of $x$ which satisfy
\begin{equation}
\dot p_{\alpha}=0,
\label{dotp}
\end{equation}
then the constraints ${\mathcal H}=0$ and ${\mathcal H}_s =0$ can be solved to express the $\varphi ^{\alpha}$ as functionals
\begin{equation}
\varphi ^{\alpha}=f^{\alpha}[\psi ^A; \pi _A]
\label{phi}
\end{equation}
of the remaining canonical variables.
\end{itemize}
The functional derivatives of $f^{\alpha}$ with respect to $\psi ^A$ and $\pi _A$ are assumed to exist. If the above conditions are true then Hamilton's equation for the reduced Hamiltonian
\begin{equation}
H^{\mathrm red}[\psi ^A; \pi _A] = ({\mathrm boundary \  terms})\bigg\vert _{\varphi ^{\alpha}=f^{\alpha}, \pi _{\alpha}=p_{\alpha}}\ ,
\label{Hred}
\end{equation}
together with Eqs. (\ref{dotp}) and (\ref{phi}) are equivalent to Einstein's equations in the particular frame defined by $\pi _{\alpha}=p_{\alpha}$.

The proof of this result is easy: The Poisson brackets are invariant under canonical transformation and the Hamiltonian is unchanged in value if the canonical transformation is independent of time. Hence
\begin{equation}
\dot \psi ^A \!(x)=\frac{\delta H}{\delta \pi _A\!(x)}\bigg\vert _{\varphi ^{\alpha}=f^{\alpha}, \pi _{\alpha}=p_{\alpha}} \ .
\label{dotpsi}
\end{equation}
On the other hand,
\begin{equation}
H[\varphi ^{\alpha}; \pi _{\alpha}, \psi ^A; \pi _A]\bigg\vert _{\varphi ^{\alpha}=f^{\alpha}, \pi _{\alpha}=p_{\alpha}}
=({\mathrm boundary \ terms})\bigg\vert _{\varphi ^{\alpha}=f^{\alpha}, \pi _{\alpha}=p_{\alpha}}=H^{\mathrm red}[\psi ^A; \pi_A]\ .
\label{Ham}
\end{equation}
Differentiating Eq. (\ref{Ham}) with respect to $\pi _A$ gives
\begin{equation}
\int _{\Sigma}d^3y \frac{\delta H}{\delta \varphi ^{\alpha}\! (y)}\bigg\vert _{\varphi ^{\alpha}=f^{\alpha}, \pi _{\alpha}=p_{\alpha}}\frac{\delta f^{\alpha}\! (y)}{\delta \pi _A \! (x)} + \frac{\delta H}{\delta \pi _A \! (x)}\bigg\vert _{\varphi ^{\alpha}=f^{\alpha}, \pi _{\alpha}=p_{\alpha}}=\frac{\delta H^{\mathrm red}}{\delta \pi _A\! (x)}\ .
\label{diff}
\end{equation}
However, by Eq. (\ref{dotp}) 
\begin{equation}
\dot \pi _{\alpha}\!(x)=\frac{\delta H}{\delta \varphi ^{\alpha}\! (x)}\bigg\vert _{\varphi ^{\alpha}=f^{\alpha}, \pi _{\alpha}=p_{\alpha}}=0\ ,
\label{dotpi}
\end{equation}
and therefore
\begin{equation}
\frac{\delta H}{\delta \pi _A\! (x)}\bigg\vert _{\varphi ^{\alpha}=f^{\alpha}, \pi _{\alpha}=p_{\alpha}}
=\frac{\delta H^{\mathrm red}}{\delta \pi _A \! (x)}\ .
\label{functder}
\end{equation}
In other words, $H^{\mathrm red}$ generates the correct equation of motion for $\psi ^A$. In a completely analogous way one shows that the correct equation of motion is also generated for $\pi _A$. Although we have here considered pure gravity only, it is clear that our analysis could be easily generalized to include electromagnetic fields as well.

The real problem is now: Are the assumptions of the previous theorem valid for Kerr-Newman spacetimes? In other words, is it possible to divide the phase space of an Einstein-Maxwell theory with appropriate symmetries in two parts in a manner explained above? For spherically symmetric, asymptotically flat Einstein-Maxwell theory this can be done and {\it has been done} in Refs.~\cite{Kuc,Wint}. For theories having the Kerr-Newman solution as their unique solution to the classical constraints this has not been done. However, there is not an obvious reason why this could not be done, and we state the following hypothesis:
\begin{itemize}
\item[]For an appropriately symmetric, asymptotically flat Einstein-Maxwell theory having the Kerr-Newman solution as its unique solution to the Hamiltonian, diffeomorphism and Gaussian constraints there exists a one to one, time independent, differentiable canonical transformation which divides the phase space $(q_{ab}, p^{ab}, A_a, p^a)$ into two sets $(M, Q, J, P_M, P_Q, P_J)$ and $(\psi ^A, P_A)$ in such a way that:
\begin{itemize}
\item[a)] The reduced Hamiltonian depends only on $M$, $Q$, $J$, $P_M$, $P_Q$ and $P_J$\ .

\item[b)] When the $M$, $Q$ and $J$ are prescribed as functions $m$, $q$, and $\iota$ which satisfy
\begin{equation}
\dot m= \dot q =\dot \iota =0\ ,
\label{dots}
\end{equation}
then the constraints can be solved to express the $P_M$, $P_Q$ and $P_J$ as functionally differentiable functionals of $\psi ^A$ and $P_A$.
\end{itemize}
\end{itemize}
We have been unable to find an exact proof of this hypothesis for Kerr-Newmann black hole spacetimes and indeed, this is the weak point of our model. However, there are no obvious reasons why it would not be true. In what follows, we shall consider our hypothesis as true and see where it takes us.  

The first consequence of our hypothesis is that $H^{\mathrm red}$ of Eq. (\ref{redHam}) may be used as the real, physical Hamiltonian of our theory, with $M$, $Q$ and $J$ as the coordinates of the configuration space. For that reason we shall drop "red" from our Hamiltonian, and denote it simply by $H$.

Our Hamiltonian now implies the following canonical equations of motion:
\begin{mathletters}
\label{eqsofmo}
\begin{eqnarray}
&\dot M&=\frac{\partial H}{\partial p_M}=0\ , 
\label{dotM}\\
&\dot Q&=\frac{\partial H}{\partial p_Q}=0\ , 
\label{dotQ}\\
&\dot J&=\frac{\partial H}{\partial p_J}=0\ , 
\label{dotJ}\\
&\dot p_M&=-\frac{\partial  H}{\partial M}=-(N^++N^-)\ ,
\label{dotpm} \\
&\dot p_Q&=-\frac{\partial H}{\partial Q}=-(A_0^+-A_0^-)\ ,
\label{dotpq} \\
&\dot p_J&=-\frac{\partial H}{\partial J}=-(\omega^+-\omega^-)\ , 
\label{dotpj}
\end{eqnarray}
\end{mathletters}
Where $p_M$, $p_Q$ and $p_J$, respectively, are canonical momenta conjugate to $M$, $Q$ and $J$.
As expected, $M$, $Q$ and $J$ are constants in time. The time derivative of $p_M$ depends on the choice of the lapse function at both asymptotic infinities of our spacelike hypersurface, $\dot p_Q$ on the difference between electric potentials at infinities, and $\dot p_J$ on the difference between the angular velocities of faraway coordinate systems.

The quantities $N^{\pm }$, $A_0^{\pm }$ and $\omega ^{\pm }$ determine the gauge of our theory. For physical reasons, it is sensible to work in a specific gauge where
\begin{mathletters}
\label{gaugefix}
\begin{eqnarray}
&N^+& \equiv 1\ ,
\label{fixN+} \\
&N^-& \equiv 0\ , 
\label{fixN-}\\
&\omega ^{\pm}& \equiv 0\ , 
\label{fixw}\\
&A_0^{\pm}& \equiv 0\ .
\label{fixA0}
\end{eqnarray}
\end{mathletters}
In this gauge the Hamiltonian takes a particularly simple form in terms of the canonical coordinates:
\begin{equation}
H=M\ .
\label{H} 
\end{equation}

The physical sense of this kind of gauge fixing lies in the fact that we consider Kerr-Newman spacetimes from the point of view of a certain specific observer: Our observer is at rest at the right hand side asymptotic infinity, and his time coordinate is the asymptotic Minkowski time, the proper time of such an observer. We have "frozen" the time evolution at the left infinity, which is sensible because our observer can make observations from just one infinity. For such an observer, the classical Hamiltonian of the Kerr-Newman spacetime is just $M$, the ADM mass of the Kerr-Newman black hole.

Now, the problem with the phase space coordinates $M$, $Q$, $J$, $p_M$, $p_Q$ and $p_J$ is that they describe the {\it static} aspects of Kerr-Newman spacetimes only. However, there is {\it dynamics} in Kerr-Newman spacetimes in the sense that between the event horizons there is a region in which it is impossible to find a timelike Killing vector field. Our next task is to find canonical variables describing the dynamical properties of Kerr-Newman black holes in a natural manner.

When choosing the phase space coordinates, we refer to the properties of our observer: Our observer lies at rest very far away from the hole and he is an inertial observer. For such an observer, the Kerr-Newman spacetime appears as stationary, and all the relevant dynamics of the Kerr-Newman spacetime is, in a certain sense, confined inside the event horizon of the hole. These properties prompt us to choose the phase space coordinates in such a manner that when the classical equations of motion are satisfied, all the dynamics is, in a certain sense, confined inside the event horizon $r=r_+$ of the hole. Moreover, as we shall see in a moment, the choice of the phase space coordinates describing the dynamics of spacetime is related to the choice of slicing of spacetime into space and time. We choose a slicing where the proper time of an observer in a free fall through the bifurcation surface coincides with the proper time of our faraway observer at rest. On grounds of the principle of equivalence one may view these types of slicings to be in a preferred position in relating the physical properties of the black hole interior to the physics observed by our faraway observer.

\subsection{Hamiltonian dynamics with charge and angular momentum as external parameters}

To make things simple, consider $J$ and $Q$ first as mere external parameters of the theory. In that case our phase space is just two-dimensional being spanned by the phase space coordinates $M$ and $p_M$. In this two-dimensional phase space we now perform the following transformation from the "old" phase space coordinates $M$ and $p_M$ to the "new" phase space coordinates $R$ and $p_R$:
\begin{mathletters}
\label{trans}
\begin{eqnarray}
&\vert p_M \vert& =\sqrt{2MR-R^2-Q^2-a^2}+M\sin^{-1}\left(\frac{M-R}{\sqrt{M^2-Q^2-a^2}}\right)+\frac{1}{2}\pi M \ , 
\label{trans1}\\ 
&p_R& = {\mathrm sgn}(p_M)\sqrt{2MR-R^2-Q^2-a^2}\ ,
\label{trans2}
\end{eqnarray}
\end{mathletters}
and we have imposed by hand a restriction
\begin{equation}
-\pi M \leq p_M \leq \pi M\ .
\label{ineq}
\end{equation}
With the restriction~(\ref{ineq}) the transformation~(\ref{trans}) is well defined and one to one. It follows from Eq.~(\ref{trans2}) that 
\begin{equation}
M=\frac{1}{2R}(p_R^2+R^2+Q^2+a^2)\ .
\label{M}
\end{equation}
If one substitutes this expression for $M$ into Eq.~(\ref{trans1}), one gets $p_M$ in terms of $R$ and $p_R$. One finds that the fundamental Poisson brackets between $M$ and  $p_M$ are preserved invariant, and hence the transformation~(\ref{trans}) is canonical.

Equations~(\ref{H}) and~(\ref{M}) imply that the classical Hamiltonian takes, in terms of the variables $R$ and $p_R$, the form
\begin{equation}
H= \frac{1}{2R}\left(p_R^2+R^2+Q^2+a^2\right)\ .
\label{HinR}
\end{equation}
The geometrical interpretation of the variable $R$ is extremely interesting. We first write the Hamiltonian equation of motion for $R$:
\begin{equation}
\dot R = \frac{\partial H}{\partial p_R}=\frac{p_R}{R}\ ,
\label{dotR}
\end{equation}
and it follows from Eq.~(\ref{HinR}) that when the classical equations of motion for $M$ and $p_M$ are satisfied, then the equation of motion for $R$ is
\begin{equation}
\dot R^2 = \frac{2M}{R}-1-\frac{Q^2+a^2}{R^2}\ .
\label{dotR'}
\end{equation}
Now, one can see from the Kerr-Newman metric~(\ref{KNelement}) that for an observer falling freely through the bifurcation surface at the equatorial plane $\theta = \pi/2$ such that  $\dot \theta = \dot \phi =0$, the proper time elapsed when $r$ goes from $r$ to $r+dr$ is $d\tau$ such that
\begin{equation}
- d\tau^2 = \frac{r^2}{r^2+a^2+Q^2-2Mr}dr^2\ ,
\label{dtau}
\end{equation}
and therefore the equation of motion of our observer is
\begin{equation}
\dot r^2 = \frac{2M}{r}-1-\frac{Q^2+a^2}{r^2}\ ,
\label{dotr}
\end{equation}
where the dot means proper time derivative. As one can see, Eqs.~(\ref{dotR'}) and~(\ref{dotr}) are identical. Hence, we may interpret $R$ as the radius of the wormhole throat of the Kerr-Newman black hole, from the point of view of an observer in a free fall at the equatorial plane such that $\dot\phi =0$ through the bifurcation two-sphere. Moreover, one can see from Eq.~(\ref{dotR'}) that $R$ is confined to be, classically, within the region $[r_-,r_+]$. In other words, our variable $R$ "lives" only within the inner and the outer horizons of the Kerr-Newman black hole, and this is precisely the region in which it is impossible to find a time coordinate such that spacetime with respect to that time coordinate would be static. Hence, both of the requirements we posed for our phase space coordinates are satisfied: Dynamics is confined inside the apparent horizon and the time coordinate on the wormhole throat is the proper time of a freely falling observer.

With the interpretation explained above, the restriction~(\ref{ineq}) becomes understandable. One can see from Eq.~(\ref{dotpm}) that when the lapse functions $N^\pm$ at asymptotic infinities are chosen as in Eqs.~(\ref{gaugefix}), the canonical momentum $p_M$ conjugate to $M$ is $-t+constant$, where $t$ is the time coordinate of our asymptotic observer. Now, the transformation (\ref{trans}) involves an identification of the time coordinate $t$ with the proper time of a freely falling observer on the throat. However, as it was noted at the beginning of this section, it is impossible to push the spacelike hypersurfaces $t=constant$ beoynd the $r=r_-$ hypersurfaces in the conformal diagram. The proper time a freely falling observer needs  to fall from the past $r=r_-$ hypersurface to the future $r=r_-$ hypersurface through the bifurcation surface is, as it can be seen from Eq.~(\ref{dotr}),
\begin{equation}
\Delta t =2\int_{r_-}^{r_+} \frac{r'dr'}{\sqrt{2Mr'-r'^2-Q^2-a^2}} = 2\pi M\ ,
\label{deltat}
\end{equation}
and hence the restriction~(\ref{ineq}) is needed. As one can see from Eq.~(\ref{trans1}), $\vert p_M\vert  =0$ when $R=r_+$ and $\vert p_M\vert  =\pi M$ when $R=r_-$. We have chosen $p_M$ to be positive when the hypersurface $t= constant$ lies between the past $r=r_-$ hypersurface and the bifurcation surface, and negative when that hypersurface lies between the bifurcation point and the future $r=r_-$ hypersurface.

\begin{figure}
    \begin{center}
      \leavevmode
      \epsfysize=90mm
      \epsfbox{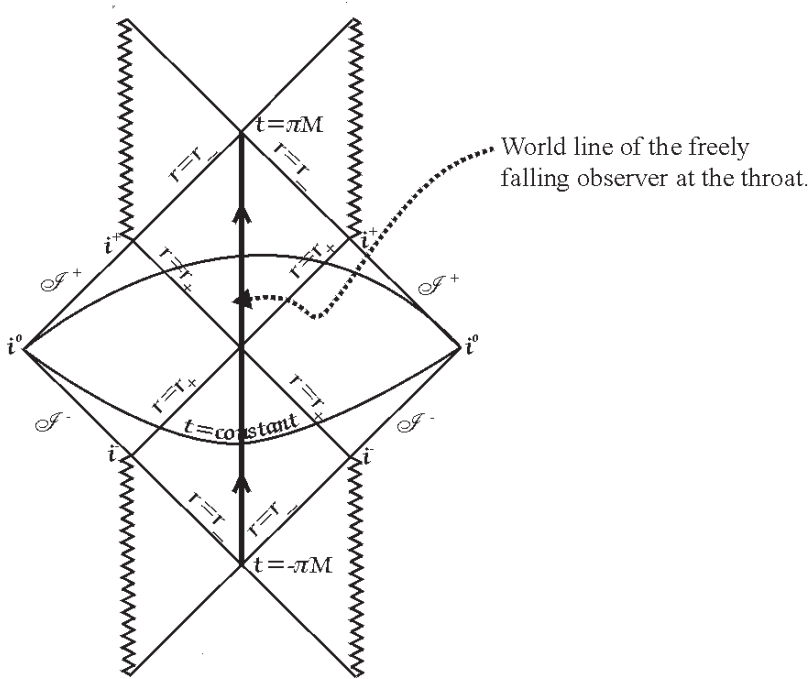}
    \end{center}
  \caption{The world line of an observer in a free fall at the throat is a vertical line going through the bifurcation point in the conformal diagram. The proper time of such an observer is identified with the asymptotic Minkowski time.}
  \label{fig2}
 \end{figure}

Concerning the classical Hamiltonian theory with $J$ and $Q$ as mere external parameters the only thing one still needs to check is whether there exists such foliations of the Kerr-Newman spacetime where the Minkowski time at asymptotic infinity  and the proper time of a freely falling observer at the throat through the bifurcation surface really are the one and the same time coordinate. It is easy to see that time coordinates determining this sort of foliations do exist. A concrete example is constructed in Appendix~\ref{app:B}. It should be noted, however, that all foliations in which the proper time on the throat and asymptotic Minkowski time are identified are incomplete since such foliations, in addition to failing to cover the regions outside the past and the future $r=r_-$ hypersurfaces also fail to cover the whole exterior regions of the hole. More precisely, these foliations are valid only when $-\pi M \leq t\leq \pi M$ (see Fig.~\ref{fig2}). 

\subsection{Hamiltonian dynamics with charge and angular momentum as dynamical variables}

The next task is to complete the classical Hamiltonian~(\ref{HinR}) such that $Q$ and $a$ are replaced by functions of appropriate phase space variables describing the dynamics of Kerr-Newman spacetimes in a natural manner. To this end, we must find, for constant $M$, a canonical transformation from the phase space coordinates $(Q, p_Q)$ and $(J, p_J)$ to some new phase space coordinates which we shall denote by $u$ and $v$, and their canonical momenta $p_u$ and $p_v$.

We shall perform such a transformation in two steps. At the first stage we replace $Q$ and $a$ by canonical momenta conjugate to yet some unknown coordinates $w_1$ and $w_2$ of the configuration space:
\begin{mathletters}
\label{pw}
\begin{eqnarray}
&p_{w_1}& := Q\ , 
\label{pw1}\\
&p_{w_2}&:=a\ ,
\label{pw2}
\end{eqnarray}
\end{mathletters} 
and the classical Hamiltonian of Eq.~(\ref{HinR}) takes the form
\begin{equation}
H= \frac{1}{2R}\left(p_R^2+p_{w_1}^2+p_{w_2}^2+R^2\right)\ .
\label{HinR'}
\end{equation}

The next task is to find $w_1$ and $w_2$. One expects that $w_1$ and $w_2$ are related in one way or another to the momenta $p_Q$ and $p_J$ conjugate to $Q$ and $J$, respectively. Because we see from Eq.~(\ref{dotpq}) that  $p_Q$ determines the electromagnetic gauge and from Eq.~(\ref{dotpj}) that $p_J$ determines the angular velocity of faraway coordinate systems we first write the classical Hamiltonian in a general electromagnetic gauge when faraway coordinate systems rotate with arbitrary angular velocities:
\begin{equation}
H= \frac{1}{2R}\left(p_R^2+p_{w_1}^2+p_{w_2}^2+R^2\right)\  + (A_0^+-A_0^-)p_{w_1}+M(\omega^+-\omega^-)p_{w_2} \ ,
\label{Hclassic}
\end{equation} 
which follows from Eq.~(\ref{redHam}). Using Eqs.~(\ref{dotpq}) and~(\ref{dotpj}) and the fact that $M$ is a constant when the classical equations of motion are satisfied, we get the Hamiltonian equations of motion for $w_1$ and $w_2$
\begin{mathletters}
\label{dotw}
\begin{eqnarray}
&\dot w_1& := \frac{\partial H}{\partial p_{w_1}} = \frac{p_{w_1}}{R}-\dot p_Q \ , 
\label{dotw1}\\
&\dot w_2& := \frac{\partial H}{\partial p_{w_2}} = \frac{p_{w_2}}{R}-M\dot p_J \ .
\label{dotw2}
\end{eqnarray}
\end{mathletters}
An expression for $p_Q$ and $p_J$ in terms of $R$, $p_R$, $w_1$, $w_2$, $p_{w_1}$ and $p_{w_2}$ can be gained by integrating both sides of Eqs.~(\ref{dotw1}) and~(\ref{dotw2}) along the classical trajectory in phase space:
\begin{mathletters}
\label{peet}
\begin{eqnarray}
&p_Q& := \int  \frac{p_{w_1}}{R\dot R}dR-w_1 \ , 
\label{pQ}\\
&p_J& := \frac{1}{M}\int  \frac{p_{w_2}}{R\dot R}dR-w_2 \ ,
\label{pJ}
\end{eqnarray}
\end{mathletters}
where we have substituted
\begin{equation}
\dot R = - {\mathrm sgn}(p_M)\sqrt{\frac{2M}{R}-1-\frac{p_{w_1}^2+p_{w_2}^2}{R}} \ ,
\label{subsR}
\end{equation}
This substitution involves choosing $\dot p_Q = \dot p_J =0$. When the electric potentials are assumed to vanish at infinities, and the asymptotic coordinate systems are assumed to be non-rotating, this kind of choice can be made. With an appropriate choice of the integration constant we get
\begin{mathletters}
\label{peet'}
\begin{eqnarray}
&p_Q& := {\mathrm sgn}(p_M) p_{w_1}\left[  \sin ^{-1} \left( \frac{p_R^2+ p_{w_1}^2+ p_{w_2}^2-R^2}{\sqrt{\left(p_R^2+ p_{w_1}^2+ p_{w_2}^2+R^2\right)^2-4R^2\left(p_{w_1}^2+ p_{w_2}^2\right) }}\right) +\frac{\pi}{2} \right] - w_1 \ , 
\label{pQ'}\\
&p_J& :=  {\mathrm sgn}(p_M) \frac{2R}{p_R^2+ p_{w_1}^2+ p_{w_2}^2-R^2}\left\lbrace  p_{w_2}\left[  \sin ^{-1} \left( \frac{p_R^2+ p_{w_1}^2+ p_{w_2}^2-R^2}{\sqrt{\left(p_R^2+ p_{w_1}^2+ p_{w_2}^2+R^2\right)^2-4R^2\left(p_{w_1}^2+ p_{w_2}^2\right) }}\right) +\frac{\pi}{2} \right] - w_2 \right\rbrace\ ,
\label{pJ'}
\end{eqnarray}
\end{mathletters}
where we have made the substitution
\begin{equation}
M = \frac{1}{2R}\left(p_R^2+p_{w_1}^2+p_{w_2}^2+R^2\right) \ .
\label{subsM}
\end{equation}

Equations~(\ref{trans2}),~(\ref{pw}) and~(\ref{peet'}) now constitute a transformation from the phase space coordinates $M$, $p_M$, $Q$, $p_Q$, $J$ and $p_J$ to the phase space coordinates $R$, $p_R$, $w_1$, $p_{w_1}$, $w_2$, and $p_{w_2}$. One can easily show that this transformation is well defined and canonical. Moreover, the transformation is one to one provided that we impose the restrictions
\begin{mathletters}
\label{restrict}
\begin{eqnarray}
&\bigg\vert \frac{p_Q+w_1}{p_{w_1}} \bigg\vert&\leq \pi \ , 
\label{resta}\\
&\bigg\vert \frac{Mp_J-w_2}{p_{w_2}} \bigg\vert&\leq \pi \ .
\label{restb}
\end{eqnarray}
\end{mathletters}
These restrictions are related to the fact that we are considering spacetime between two successive $r=r_-$ hypersurfaces. Since both $\dot p_Q$ and $\dot p_J$ vanish when the electric potentials are assumed to vanish at asymptotic infinities and the asymptotic coordinate systems are assumed to be non-rotating, we find that classically $w_1$ and $w_2$ have the following properties: At the past $r=r_-$ hypersurface $w_1=-Q\pi+p_Q$ and $w_2 = -a\pi+Mp_J$, at the bifurcation surface $w_1=p_Q$ and $w_2=Mp_J$, and at the future $r=r_-$ hypersurface $w_1 = Q\pi+p_Q$, and $w_2 =a\pi +Mp_J$. In other words, the classical domains of $w_1$ and $w_2$ are bounded by the fact that  the $t= constant$ hypersurfaces cannot be pushed beyond the $r=r_-$ hypersurfaces.

As the last step we perform a canonical transformation from the variables $w_1$, $p_{w_1}$, $w_2$  and $p_{w_2}$ to the variables $u$, $p_u$, $v$ and $p_v$\footnote{$u$ and $v$ should not be confused with light cone coordinates or anything like that!}. We define
\begin{mathletters}
\label{last}
\begin{eqnarray}
&u&:=p_{w_1}\sin\left( \frac{w_1}{p_{w_1}}\right)\ , 
\label{u}\\
&p_u&:=p_{w_1}\cos\left( \frac{w_1}{p_{w_1}}\right) \ ,
\label{pu}\\
&v&:=p_{w_2}\sin\left( \frac{w_2}{p_{w_2}}\right)\ ,
\label{v}\\
&p_v&:=p_{w_2}\cos\left( \frac{w_2}{p_{w_2}}\right)\ .
\label{pv}
\end{eqnarray}
\end{mathletters}
This transformation is well defined, canonical and, with the restriction~(\ref{restrict}), one to one as well. We find that
\begin{mathletters}
\label{pwpw}
\begin{eqnarray}
&p_{w_1}^2&=p_u^2+u^2\ ,
\label{pw1pw1}\\
&p_{w_2}^2&:=p_v^2+v^2\ .
\label{pw2pw2}
\end{eqnarray}
\end{mathletters}
In other words, we may identify $p_u^2+u^2$ as the square $Q^2$ of the electric charge $Q$, and $p_v^2+v^2$ as the square of $a^2$ of the angular momentum per unit mass of the hole. Because of that, the classical Hamiltonian of Kerr-Newman black holes finally takes a very simple form
\begin{equation}
H=\frac{1}{2R}\left(p_R^2+p_u^2+p_v^2+R^2+u^2+v^2 \right) \ .
\label{Hbeauty}
\end{equation}

\section{Quantum theory of Kerr-Newman black holes}
\label{sec:Quant}

After completing the classical Hamiltonian theory of stationary spacetime containing a Kerr-Newman black hole, we are now prepared to consider the canonical quantum theory of such spacetimes. In what follows, we shall concentrate on a specific class of canonical quantum theories. More precisely, we define the Hilbert space to be the space $L^2({\mathbb R}^+ \times {\mathbb R} \times {\mathbb R}, R^sdRdudv)$ with the inner product
\begin{equation}
\langle\Psi_1\vert \Psi _2\rangle:=\int _0^{\infty} dRR^s\int _{-\infty}^{\infty}du\int _{-\infty}^{\infty}dv \Psi_1^*\!(R, u, v)\Psi _2\!(R, u, v)\ .
\label{inner}
\end{equation}
Through the substitutions $p_R \rightarrow -i\frac{\partial}{\partial R}, p_u \rightarrow -i\frac{\partial}{\partial u}$ and $p_v \rightarrow -\frac{\partial}{\partial v}$ we replace the classical Hamiltonian of Eq.~(\ref{Hbeauty}) by the corresponding symmetric operator
\begin{equation}
\hat H := -\frac{1}{2}R^{-s}\frac{\partial}{\partial R}\left(R^{s-1}\frac{\partial}{\partial R}\right) - \frac{1}{2R}\frac{\partial ^2}{\partial u^2}-\frac{1}{2R}\frac{\partial ^2}{\partial v^2}+\frac{1}{2}R+\frac{u^2}{2R}+\frac{v^2}{2R} \ .
\label{Hsymm}
\end{equation}
This operator may be viewed as the Hamiltonian operator of Kerr-Newman black holes. Its eigenvalues are eigenvalues of the ADM energy $E$ of such a hole, from the point of view of a faraway observer at rest. The eigenvalue equation in question takes the form:
\begin{equation}
\left[-\frac{1}{2}R^{-s}\frac{\partial}{\partial R}\left(R^{s-1}\frac{\partial}{\partial R}\right) - \frac{1}{2R}\frac{\partial ^2}{\partial u^2}-\frac{1}{2R}\frac{\partial ^2}{\partial v^2}+\frac{1}{2}R+\frac{u^2}{2R}+\frac{v^2}{2R}\right]\Psi\! (R,u,v) = E\Psi \!(r,u,v) \ .
\label{eigeneq}
\end{equation}
This equation is the main result of this paper. In a certain sense, it can be considered as a sort of a "time-independent Schr\"odinger equation of all the possible black holes", and $\Psi \! (R,u,v)$ as the wave function of black holes. Concentrating on the quantum theories where $s=1$, we find that Eq.~(\ref{eigeneq}) takes a particularly simple and beautiful form:
\begin{equation}
\frac{1}{2R}\left(-\frac{\partial ^2}{\partial R^2}-\frac{\partial ^2}{\partial u^2}-\frac{\partial ^2}{\partial v^2}+R^2+u^2+v^2\right)\Psi \!(R,u,v) = E\Psi \! (R,u,v) \ .
\label{eigenbeauty}
\end{equation}
If we write
\begin{equation}
\Psi\! (R,u,v)=\psi \!(R)\varphi _1\! (u)\varphi _2\!(v) \ , 
\label{separate}
\end{equation}
we find that Eq.~(\ref{eigeneq}) can be separated to eigenvalue equations for $M$, $Q^2$ and $a^2$
\begin{mathletters}
\label{eigeqs}
\begin{eqnarray}
\left[-\frac{1}{2}R^{-s}\frac{d}{dR}\left(R^{s-1}\frac{d}{dR}\right) +\frac{1}{2}R+\frac{Q^2}{2R}+\frac{a^2}{2R}\right]\psi \!(R)&=&M\psi \!(R) \ , \label{eigeqsa} \\
\left(-\frac{d^2}{du^2}+u^2 \right)\varphi _1 \!(u) &=& Q^2\varphi_1 \!(u) \ , \label{eigeqsb} \\
\left(-\frac{d^2}{dv^2}+v^2 \right)\varphi _2 \!(v) &=& a^2\varphi_2 \!(v) \ . \label{eigeqsc} 
\end{eqnarray}
\end{mathletters}

Consider now Eq.~(\ref{eigeqsa}), the eigenvalue equation for the ADM mass $M$ of the hole, in more details. It can be written as:
\begin{equation}
\left[R^{-s}\frac{d}{dR}\left(R^{s-1}\frac{d}{dR}\right)\right]\psi \!(R)= \left(\frac{Q^2}{R}+\frac{a^2}{R}+R-2M\right)\psi \!(R) \ . 
\label{ADMMeq}
\end{equation}
As one can see, the function
$$
\frac{Q^2}{R}+\frac{a^2}{R}+R-2M
$$
is negative when $r_-<R<r_+$ and positive (or zero) elsewhere. Semiclassically, one may therefore expect oscillating behavior from the wave function when $r_-<R<r_+$ and exponential behavior elsewhere. Hence, our system is somewhat analogous to a particle in a potential well such that $R$ is confined, classically, between the outer and the inner horizons of the black hole. What happens semiclassically is that the wave packet corresponding to the variable $R$ is reflected from the inner  horizon. As a result, we get, when the hole is in a stationary state, a standing wave between the outer and the inner horizons. Thus the classical incompleteness, associated with the fact that our foliation is valid only when $-\pi M\leq t\leq \pi M$, is removed by quantum mechanics: In a stationary state there are no propagating wave packets between the horizons and our quantum theory is therefore valid at any moment of time.

When $a=Q=0$, we have a Schwarzschild black hole, and the inner horizon is replaced by the black hole singularity: The wave packets are no more reflected from the inner horizon but from the singularity. Again, we have a standing wave in a stationary state and the quantum theory is valid at any moment of time, but the wave lies between Schwarzschild horizon and the singularity. As such there is an interesting resemblance between the properties of Eq.~(\ref{eigeneq}) and those of the Schr\"odinger equation of a hydrogen atom: When the hydrogen atom is in an s-state where the orbital angular momentum of the electron orbiting the proton vanishes, the electron should, classically, collide with the proton in a very short time. Quantum mechanically, however, the wave packet representing the electron is reflected from the proton, and finally the electron is represented by a standing wave, which makes the quantum theory of the hydrogen atom valid at any moment of time. In a Schwarzschild black hole, the proton is replaced by the black hole singularity, and the distance of an electron from the proton is replaced by the throat radius $R$ of the hole. Nevertheless, the solution provided by quantum theory to the problems of the classical one is similar for both black holes and hydrogen atoms.

We shall now enter the detailed analysis of the eigenvalue equation~(\ref{eigeqsa}). To begin with, we see that if we denote
\begin{mathletters}
\label{denote}
\begin{eqnarray}
&x&:=R^{3/2} \ , \label{denotea} \\
&\psi& := x^{(1-2s)/6}\chi\!(x)\ , \label{denoteb} 
\end{eqnarray}
\end{mathletters}
and define
\begin{mathletters}
\label{def}
\begin{eqnarray}
&\rho&:=\frac{2s-1}{6}\  , s \geq 2  \ , \label{defa} \\
&\rho& := \frac{7-2s}{6}\ , s < 2 \label{defb} 
\end{eqnarray}
\end{mathletters} 
then Eq.~(\ref{eigeqsa}) takes the form
\begin{equation}
\frac{9}{8}\left[-\frac{d^2}{dx^2} +\frac{\rho(\rho-1)}{x^2}+\frac{4}{9}\left( x^{2/3}+\frac{Q^2+a^2}{x^{2/3}}\right)\right]\chi \!(x) = M\chi\!(x) \ . \label{Meigeq} 
\end{equation}
This equation has been analyzed in details in Ref.~\cite{Mak2}. The only difference between Eq.~(\ref{Meigeq}) and Eq.~(3.18) in Ref.~\cite{Mak2} is that $Q^2$ has been replaced by $Q^2+a^2$. Hence one just replaces $Q^2$ by  $Q^2+a^2$ in the results obtained for Eq.~(3.18) in Ref.~\cite{Mak2}.

As in Ref.~\cite{Louko}, one can show that the spectrum of $M$ is discrete, bounded below, and can be made positive. From the physical point of view, the semiboundedness and positivity (in some cases) of the spectrum are very satisfying results: The semiboundedness of the spectrum implies that one cannot extract an infinite amount of energy from the system, whereas the positivity of the spectrum is in agreement with the well-known positive-energy theorems of general relativity, which state, roughly speaking, that the ADM energy of spacetime is always positive or zero when Einstein's equations are satisfied~\cite{Wald}.

However, one can prove even more than that. One can show that the eigenvalue equation~
(\ref{Meigeq}) implies that when $\rho \geq 3/2$, the eigenvalues of the quantity
$$
M^2-Q^2-a^2
$$
are strictly positive, and when $1/2 \leq \rho < 3/2$, the eigenvalues of the quantity $M^2-Q^2-a^2$ can again be made positive by means of an appropriate choice of the boundary conditions of the wave function $\chi\!(x)$ at the point $x=0$ or, more precisely, by means of an appropriate choice of a self-adjoint extension. Moreover, the WKB analysis of Eq.~(\ref{Meigeq}) yields the result that when $Q^2+a^2 \gg 1$, and $M^2-Q^2-a^2 \gg 1$ such that $r \gg 1$, the WKB eigenvalues $M_n$ have a property
\begin{equation}
M_n^2-Q^2-a^2 \sim 2n+1 +{\cal O}(1) \ ,
\label{valM}
\end{equation}
where $n$ is an integer and ${\cal O}(1)$ denotes a term that vanishes asymptotically for large $n$. A numerical analysis of  Eq.~(\ref{Meigeq}) yields the result that, up to the term 1 on the right hand side, Eq.~(\ref{valM}) gives fairly accurate results even when $\sqrt{Q^2+a^2}$ and $n$ are relatively small (i.e., of order 10). In other words, it seems that the eigenvalues of the quantity $\sqrt{M^2-Q^2-a^2}$ are of the form $\sqrt{2n}$ in the semiclassical limit.

Now, how should we understand these results? The positivity of the spectrum of the quantity $M^2-Q^2-a^2$ has an interesting consequence regarding Hawking radiation: If one thinks of Hawking radiation as an outcome of a chain of transitions from higher- to lower-energy eigenstates, the positivity of the spectrum of $M^2-Q^2-a^2$ implies that a non-extreme Kerr-Newman black hole can never become, through Hawking radiation, an extreme black hole with zero temperature, a result that is in agreement with both the third law of thermodynamics and the qualitative difference between extreme and non-extreme black holes. One may consider this as a strong argument in favor of our choice of the phase space coordinates describing the dynamics of Kerr-Newman spacetimes.

Before considering the implications of Eq.~(\ref{valM}), let us calculate the spectra of $Q$ and $a$ from Eqs.~(\ref{eigeqsb}) and~(\ref{eigeqsc}). As one can see, both of these equations are, essentially, time-independent Schr\"odinger equations of a one-dimensional linear harmonic oscillator. When the solutions to Eq.~(\ref{eigeqsb}) are chosen to be harmonic oscillator eigenfunctions, one finds that the eigenvalues of $Q^2$ are
\begin{equation}
Q^2_k=2k+1\ ,
\label{valQ}
\end{equation}
or, in SI units,
\begin{equation}
Q^2_k=(2k+1)\frac{e^2}{\alpha}\ ,
\label{valQ'}
\end{equation} 
where $k= 0,1,2,\dots$. In this equation, $e$ is the elementary charge and
\begin{equation}
\alpha :=\frac{e^2}{4\pi\epsilon_0\hbar c}\approx\frac{1}{137}
\label{fine}
\end{equation}
is the fine structure constant. In other words, Eq.~(\ref{eigeneq}), the "Shr\"odinger equation of black holes", implies that the electric charge of black holes has a discrete spectrum.

One may have very mixed feelings on the physical validity of the charge spectrum in Eq.~(\ref{valQ'}): For elementary particles at least, the electric charge $Q$ itself, instead of its square $Q^2$ is an integer. Because of that it might appear at the first sight that the charge spectrum we have just obtained contradicts all the possible observations and expectations, and should therefore be rejected on physical grounds.

Such a conclusion, however, would be much too rapid. Firstly, elementary particles are certainly not black holes because for them $\vert Q\vert \gg M$. Secondly, a dimensional investigation reveals us that the charge spectrum (\ref{valQ}) is exactly what one expects for black holes: when one writes the electric charge in terms of the natural constants $\epsilon_0$, $\hbar$ and $c$, one finds that the natural unit of electric charge is the so called "Planck charge"
\begin{equation}
Q_{\mathrm Pl}:=\sqrt{4\pi\epsilon_0\hbar c}\ .
\label{Plcharge}
\end{equation}
One observes that the square $Q_{\mathrm Pl}^2$ of the Planck charge $Q_{\mathrm Pl}$, instead of the Planck charge $Q_{\mathrm Pl}$ itself, is proportional to $\hbar$. Now, for bounded systems, the observed quantities usually tend to be quantized in such a manner that when we write that quantity in terms of the natural constants relevant to the system under consideration, then $\hbar$ must be multiplied by an integer in the spectrum. In a hydrogen atom, for instance, the relevant natural constants are $\epsilon_0$, $\hbar$, $e$ and the mass $m_e$ of the electron. From these quantities one may construct a natural unit of energy in a hydrogen atom:
$$
\frac{m_ee^4}{(4\pi\epsilon_0)^2\hbar^2}\ ,
$$
and one expects the energy to be quantized such that the energy eigenvalues are of the form
\begin{equation}
E_n=-\gamma \frac{m_ee^4}{(4\pi\epsilon_0)^2\hbar^2n^2}\ ,
\label{hydro}
\end{equation}
where $\gamma$ is some pure number and  $n$ is an integer. Indeed, if we take $\gamma = 1/2$, we get exactly the correct energy spectrum for a hydrogen atom. Now, for black holes the only natural constants we are allowed to use are, in SI units, $\hbar$, $c$, $G$ and $\epsilon_0$. Hence, the Planck charge $Q_{\mathrm Pl}$ of Eq.~(\ref{Plcharge}) is the natural unit of charge for black holes, and therefore one expects that the square of the electric charge, instead of the charge itself, must be an integer. In other words, the charge spectrum~(\ref{valQ'}) is exactly what one expects for black holes.

Let us now turn our attention to Eq.~(\ref{eigeqsc}) which gives the spectrum of $a^2$. As for the electric charge, we find that the possible eigenvalues of $a^2$ are
\begin{equation}
a_l^2=2l+1
\label{vala}
\end{equation}
or, in SI units,
\begin{equation}
a_l^2=(2l+1)\frac{\hbar G}{c}\ ,
\label{vala'}
\end{equation}
where $l=0,1,2,\dots$. Again, one observes that the quantity under consideration is quantized in such a way that $\hbar$ is multiplied by an integer. Putting Eqs.~(\ref{valM}),~(\ref{valQ'}) and~(\ref{vala}) together we find that, semiclassically, the mass eigenvalues of the black hole are
\begin{equation}
M_m \sim \sqrt{2m}
\label{Mm}
\end{equation}
or, in SI units
\begin{equation}
M_m\sim \sqrt{2m}M_{\mathrm Pl} \ ,
\label{Mm'}
\end{equation}
where
\begin{equation}
m:=n+l+k=0,1,2,\dots\ ,
\label{m}
\end{equation}
and
\begin{equation}
M_{\mathrm Pl}:=\sqrt{\frac{\hbar c}{G}}
\label{MPl}
\end{equation}
is the Planck mass.

The spectra of the quantities $M$, $Q$ and $a$ now have interesting consequences regarding the area spectrum of black holes. As it is well known, the area of the outer horizon of a Kerr-Newman black hole is
\begin{equation}
A^+=4\pi (r_+^2 + a^2) \ ,
\label{A+}
\end{equation}
whereas the area of the inner horizon is
\begin{equation}
A^-=4\pi (r_-^2 + a^2) \ .
\label{A-}
\end{equation}
Using Eqs.~(\ref{valM}) and~(\ref{valQ'}) we observe that the semiclassical eigenvalues of the quantity
\begin{equation}
A^{\mathrm tot}:=A_+ + A_- \ ,
\label{Atot}
\end{equation}
which we shall call, for the sake of convenience, the {\it total area} of a black hole, are of the form
\begin{equation}
A^{\mathrm tot}_{n,l,k}\sim 16\pi (2n + 2l + k)
\label{Atotnlk}
\end{equation}
or, in SI units,
\begin{equation}
A^{\mathrm tot}_{n,l,k}\sim 16\pi (2n + 2l + k)l^2_{\mathrm Pl}\ ,
\label{Atotnlk'}
\end{equation}
where
\begin{equation}
l_{\mathrm Pl}:=\sqrt{\frac{\hbar G}{c^3}}
\label{Planck}
\end{equation}
is the Planck length. In other words, we have obtained a result which is closely related, although not quite identical to the proposal suggested by Bekenstein in 1974 and since then revived by several authors:  According to that proposal the spectrum of the outer horizon of the black hole is of the form~\cite{Bek1,Muk,Bek2,Bek3,Maz,Kog,Mag,Lousto,Pel,Louko,Bar,Kas1,Mak1,Hod1,Mak2,Hod2,Hod3,Kas2,Bojo,Ah,Gar}
\begin{equation}
A^+_n=\gamma nl_{\mathrm Pl}^2 \ ,
\label{A+n}
\end{equation}
where $n$ is integer and $\gamma$ is a pure number of order one. Hence, we have obtained a result which states that the total area of the hole, with $\gamma = 16\pi$, instead of the area of its inner horizon, is quantized as in Eq.~(\ref{A+n}). In the last section of this paper we shall consider in more details the possibility that it is perhaps the total area, and not the area of the outer horizon, which should be an integer in Planck units.

As the final check of our quantum theory of black holes, let us calculate the angular momentum spectrum of black holes. We observe from Eqs.~(\ref{vala'}),~(\ref{Mm'}) and~(\ref{m}) that the possible eigenvalues of the angular momentum $J=Ma$ of the hole are, semiclassically, of the form
\begin{equation}
J_{n,l,k} \sim \pm 2\sqrt{l(l+n+k)}\hbar \ .
\label{Jnlk}
\end{equation}
For uncharged black hole where $k=0$ we therefore find, in the limit of extremality where $l\gg n$, that the angular momentum eigenvalues are of the form
\begin{equation}
J_{m_j}\sim m_j\hbar \ ,
\label{Jmj}
\end{equation}
where $m_j=0, \pm 2, \pm 4, \dots$.

As one can see, the angular momentum spectrum of black holes, as predicted by our theory is, at least in the limit of extremality, exactly what one might expect. Even the fact that the angular momentum $J$ is an even number, is in harmony with our expectations: When the black hole performs a transition from one angular momentum eigenstate to another, a graviton is emitted or absorbed. Because the spin of a graviton is two, one might expect that the angular momentum of the black hole could change only by an even number. For instance, one may show, quite rigorously, that when a system consisting of two mass points revolving around their common center of mass emits or absorbs a graviton, the angular momentum quantum number of the system can change only by an even number~\cite{Repo}. Because of that, the angular momentum spectrum given by Eq.~(\ref{Jmj}) for extremal black holes may be used as a very strong argument in favor of the physical validity of our quantum-mechanical model of black holes.

Unfortunately, our model also appears to contain a very serious problem regarding the angular momentum spectrum: According to Eq.~(\ref{Jnlk}) the angular momentum of a black hole is not in general an integer times the Planck constant $\hbar$. Should we be worried because of this result?

The answer to this question is: Not necessarily. The usual rules for the quantum mechanics of angular momentum follow from the symmetries of {\it flat spacetime}, and spacetime containing a Kerr-Newman black hole is certainly not flat. In curved spacetime the angular momentum eigenvalues of a system do not necessarily have the same properties as they would have in flat spacetime.

\begin{figure}
    \begin{center}
      \leavevmode
      \epsfysize=60mm
      \epsfbox{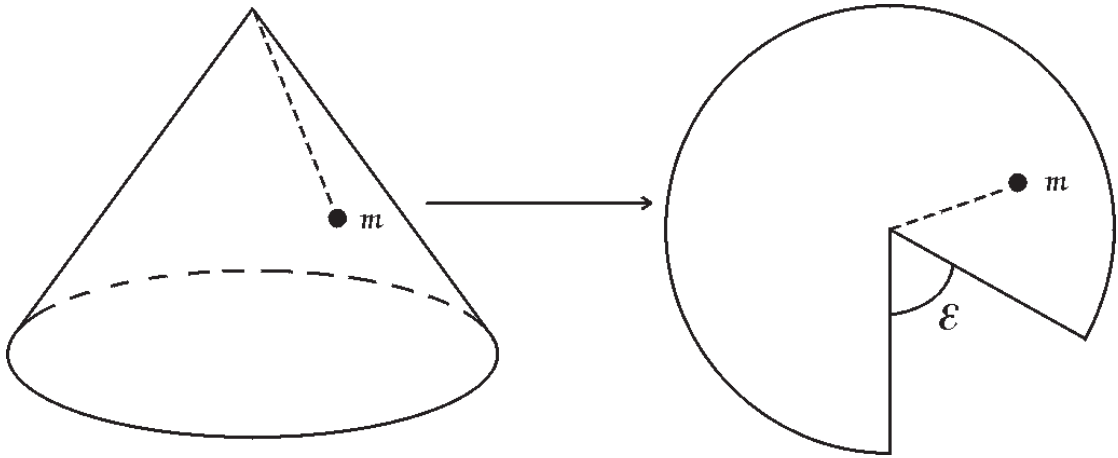}
    \end{center}
  \caption{A particle moving in a conelike spacetime geometry. When the cone is stretched on a plane, the deficit angle $\varepsilon$ appears. As a result of the appearance of this deficit angle, the periodic boundary condition for the angular momentum eigenfunction $\psi (\phi )$ is $\psi (\phi+2\pi-\varepsilon ) = \psi (\phi)$.}
  \label{fig3}
 \end{figure}

To illustrate this fact by a simple example, consider a particle moving in a conelike spacetime geometry (See Fig.~3). The $z$-component $L_z$ of the angular momentum eigenvalues may be calculated from the equation
\begin{equation}
-i\hbar \frac{\partial}{\partial \phi}\Psi \!(\phi)=L_z\Psi \! (\phi) \ ,
\label{Lz}
\end{equation}
from which it follows that the angular momentum eigenfunctions are of the form
\begin{equation}
\Psi \! (\phi)=Ce^{\frac{i}{\hbar}L_z\phi} \ ,
\label{eigenfunct}
\end{equation}
where $C$ is a constant. In flat spacetime the period of $\Psi \! (\phi)$ is $2\pi$, producing the usual angular momentum spectrum. In conelike spacetime geometry, however, the period of $\Psi$ is {\it not} $2\pi$ but $2\pi - \epsilon$, where $\epsilon$ is the deficit angle of the cone (See Fig.~\ref{fig3}). In other words, we must have:
\begin{equation}
\Psi \! (\phi +2\pi -\epsilon)=\Psi \! (\phi) \ ,
\label{epsilon}
\end{equation}
and therefore the angular momentum eigenvalues are of the form
\begin{equation}
L_z=m_z\frac{1}{1-\frac{\epsilon}{2\pi}}\hbar \ ,
\label{Lze}
\end{equation}
where $m_z=0,\pm 1, \pm 2, \dots$. In other words, the angular momentum of a system is not necessarily an integer times the Planck constant in curved spacetime.

\section{Concluding remarks}
\label{sec:Conc}

In this paper we have considered a particular quantum-mechanical model of Kerr-Newman black holes. The fundamental ideas behind our model were based on the black hole uniqueness theorems. According to these theorems a black hole in stationary spacetime is completely characterized by exactly three free variables which may taken to be the mass $M$, the electric charge $Q$ and the angular momentum $J$ of the hole. From these theorems it follows that the Kerr-Newman solution, being completely characterized by these three free variables, is the most general stationary black hole solution to combined Einstein-Maxwell equations. In our model we considered a Hamiltonian quantum theory of stationary black hole spacetimes in such a way that the phase space was spanned by the variables $M$, $J$ and $Q$, together with the corresponding canonical momenta $p_M$, $p_J$ and $p_Q$. The problem with these phase space coordinates, however, is that they describe the {\it static} aspects of black hole spacetimes only. However, there is {\it dynamics} in Kerr-Newman spacetimes in the sense that between the horizons there is no timelike Killing vector field, and we managed to find new phase space coordinates which describe the dynamical properties of Kerr-Newman spacetime in a particularly natural manner. These phase space coordinates were replaced by the corresponding quantum mechanical operators yielding the symmetric Hamiltonian operator. Our analysis produced Eq.~(\ref{eigeneq}) which, in a certain very restricted sense, may be considered as a sort of "Schr\"odinger equation of black holes". That equation gives, in the context of our model, the mass, electric charge and angular momentum spectra of black holes.

Eq.~(\ref{eigeneq}), which is the main result of this paper, implies that the mass, electric charge and angular momentum spectra of black holes are discrete. Moreover, it implies that the mass spectrum is bounded from below and can be made positive. By means of a choice of an appropriate self-adjoint extension one may show that the spectrum of the quantity
$$
M^2-Q^2-a^2 \ ,
$$
where $a$ is the angular momentum per unit mass of the hole, is always positive. Regarding Hawking radiation, this is a very important result: It means that a non-extreme black hole can never become an extreme black hole by means of the Hawking radiation of black holes. This result is in agreement with the third law of black hole thermodynamics, and is therefore a strong argument in favor of the physical validity of our model.

At the high end of the spectrum, Eq.~(\ref{eigeneq}) implied that the eigenvalues of the quantities $M$, $Q$ and $a$ are all quantized, in natural units, in a very similar manner: In natural units the eigenvalues of these quantities are all of the form $\sqrt{2n}$, where n is an integer. Although this kind of a spectrum might appear very odd for an electric charge spectrum of black holes, it is exactly what one expects on dimensional grounds. In the extremal limit, Eq.~(\ref{eigeneq}) implied that the angular momentum eigenvalues of black holes are of the form
$$
m_j\hbar \ ,
$$
where $m_j=0,\pm 2, \pm 4, \dots $.

Of particular interest is the area spectrum of black holes given by Eq.~(\ref{eigeneq}). Eq.~(\ref{eigeneq}) implied that the sum of the areas of the two horizons of Kerr-Newman black hole is of the form
$$
n16\pi l^2_{\mathrm Pl} \ ,
$$
where $l_{\mathrm Pl}$ is the Planck length. Hence, we get a result which is closely related, although not quite identical to, the proposal made by Bekenstein in 1974. According to Bekenstein's proposal, it is not the sum of areas of horizon but the area of the outer horizon which has an equal spacing in its spectrum.

Although our result about an equal spacing for the spectrum of the sum of the horizon areas may have certain esthetic merits, it also involves some problems. For instance, the fact that the mass eigenvalues are of the form $\sqrt{2m}$ which, together with the fact that $Q$ and $a$ have similar spectra, implied the area spectrum under consideration, also implies that the angular frequencies of quanta of Hawking radiation emitted in transitions between nearby states is
\begin{equation}
\omega \approx \frac{1}{M}\ .
\label{omegam}
\end{equation}
For Schwarzschild black holes this is something one might expect because the Hawking temperature of such a hole is~\cite{Hart}
\begin{equation}
T_{\mathrm H}=\frac{1}{8\pi M} \ ,
\label{Htemp}
\end{equation}
and therefore it follows from Wien's displacement law that the maximum of the thermal spectrum of black hole radiation corresponds to the angular frequency
\begin{equation}
\omega _{\mathrm max} \propto \frac{1}{M} \ .
\label{omegamaxm}
\end{equation}
In other words, the angular frequency associated with the discrete spectrum of Hawking radiation as predicted by our model, behaves, as a function of $M$, in the same way as does the angular frequency corresponding to the maximum of the thermal spectrum as predicted by Hawking and others.

Unfortunately, this nice correspondence between Hawking's result and our model breaks down when $Q$ or $a$ are different from zero. In that case the Hawking temperature of the black hole is~\cite{Hart}
\begin{equation}
T_{\mathrm H}=\frac{\sqrt{M^2-Q^2-a^2}}{2\pi \left[\left(M+\sqrt{M^2-Q^2-a^2}\right)^2+a^2\right] } \ ,
\label{Th}
\end{equation}
and one finds that the maximum of the thermal spectrum corresponds to the angular frequency
\begin{equation}
\omega _{\mathrm max}\propto \frac{\sqrt{M^2-Q^2-a^2}}{\left(M+\sqrt{M^2-Q^2-a^2}\right)^2+a^2 } \ .
\label{omegamax}
\end{equation}
In other words, the angular frequency (\ref{omegam}) predicted by our model corresponds, when the hole is near extremality, to a temperature which is much {\it higher} than the Hawking temperature. 

However, there may be a possible way out of this problem. In all our investigation we have emphasized the importance of the dynamics of the intermediate region between the horizons of the black hole. The dynamics of this intermediate region is, in our model, responsible for the discrete eigenvalues of the mass, electric charge and angular momentum of the hole. Now, if we take this point of view to its extreme limits we are tempted to speculate that both of the horizons of the hole, acting as the boundaries of the intermediate region, may participate, in one way or another, in the radiation process of the black hole. In other words, both of the horizons may radiate. The radiation emitted by the inner horizon is probably emitted inside the inner  horizon, and is therefore not observed by the external observer. Nevertheless, an emission of this radiation is likely to reduce considerably the number of quanta, and hence the temperature, of the radiation coming out from the hole: The more the inner horizon radiates, the less quanta are left for the outer horizon.

Let us give up for a moment our resistance to this most charming temptation and have a play with the thought that both of the horizons have an important role in black hole radiation. For instance, one might consider one quarter of the total area of the hole as a sort of a "total entropy" of the hole
\begin{equation}
S_{\mathrm tot}=\frac{1}{4}(A^+ + A^-) \ .
\label{totalS}
\end{equation}
Moreover, one might be inclined to define a temperature $T$ corresponding to this entropy (whatever that means) such that
\begin{equation}
\frac{1}{T}:=\frac{\partial S_{\mathrm tot}}{\partial E} \ ,
\label{totalST}
\end{equation}
and one finds, quite unexpectedly, that
\begin{equation}
T=\frac{1}{8\pi M} \ .
\label{TpiM}
\end{equation}
In other words, we have recovered the Hawking temperature of the Schwarzschild black hole (see Eq.~(\ref{Htemp})). This expression is the same for all black holes, and it is inversely proportional to the mass $M$ of the hole. It may well be that all this is just meaningless play with symbols, without any physical content, but nevertheless the idea that it is the total area, and not the area of the outer horizon, which is of fundamental importance in black hole quantum mechanics, appears to possess remarkable internal consistency: If the total area of the hole has equal spacing in its spectrum, one expects the temperature of the hole to be inversely proportional to the mass $M$, and this result is recovered if the total entropy of the hole is taken to be one quarter of not the area of the outer horizon but of the total area of the hole. We shall investigate these ideas in more details in forthcoming papers.

To conclude, our quantum-mechanical model of Kerr-Newman black holes appears to involve several physically sensible properties but also some problems. For instance, the claim that Kerr-Newman spacetime and our phase space variables satisfy the assumptions of Regge's and Teitelboim's theorem has been left unproved. The proper analysis of the Hamiltonian dynamics of Kerr-Newman spacetimes along the lines shown by Kucha\v r for Schwarzscild spacetime should therefore be performed~\cite{Kuc}.

Another problem is, whether the quantum mechanics of black holes can be described with a sufficient accuracy by means of a model having just three independent degrees of freedom. In other words, are the mass, electric charge and angular momentum spectra obtained from our model reliable? When answering to this question one can just say that at least the spectra are such as one might expect on semiclassical and dimensional grounds. As to the problems related to the statistical origin of black hole entropy and things like that our model says nothing. Another, more esoteric, reason why our model may probably contain some hints of truth is its simplicity and certain naturality. Such things, however, are merely matters of taste and should not be trusted too much.

\acknowledgments

We are grateful to Markku Lehto and Jorma Louko for useful discussions and constructive criticism during the preparation of this paper. 

\appendix

\section{A Boundary term}
\label{app:A}

In this Appendix the boundary term $S^{\mathrm rev}_{\partial \Sigma}$ of 
Section~\ref{sec:Bound} is calculated explicitly. 

That boundary term is expressed in Cartesian coordinates in Eq.~(\ref{boundrot}). 
Because the coordinate system revolves around the z-axis with angular velocity 
$\omega$, we have $\omega^1 = 0 = \omega^2$ 
and $\omega^3 = \omega$, and so the boundary term can be written in the form
\begin{equation}
S^{\mathrm rev}_{\partial \Sigma}=-2 \int dt  \; \omega \oint \left( x p ^2 _{\ s} dS^s - y p ^1 
_{\ s} dS^s \right) \ . 
\end{equation}

By comparing the line element of Eq.~(\ref{appKNelement'}) with the ADM line 
element
\begin{equation}
ds^2= - \left(N^2-N_aN^a\right)dt^2 + 2N_adx^adt + q_{ab}dx^adx^b \ ,
\label{appADMelement}
\end{equation}
where $N$ is the lapse function, $N^a$ is a component of the shift vector (a = 1,2,3) 
and $q_{ab}$ is a spacelike component of 
the metric tensor associated with the hypersurface,
we find that the only non-zero spacelike components of the metric tensor are
\begin{equation}
q_{11} = 1+\frac{2M}{r} = q_{22} = q_{33}
\label{appmetric}
\end{equation}
and for the components of the shift vector $N^a$ we have
\begin{equation}
N_1 = \frac{2Jy}{r^3}, \ N_2 = -\frac{2Jx}{r^3}, \ N_3 = 0
\end{equation}
where
\begin{equation}
N_a :=q_{ab}N^b
\end{equation}
and
\begin{equation}
r=\sqrt{x^2+y^2+z^2} \ .
\end{equation}

Comparing Eqs. ~(\ref{appKNelement'}) and ~(\ref{appADMelement}) we obtain the 
lapse function N:
\begin{equation}
N=\sqrt{1-\frac{2M}{r}+\frac{4J^2}{r^6}\left(1+\frac{2M}{r}\right)^{-
1}\left(x^2+y^2\right)} \ .
\end{equation}

To evaluate covariant derivatives of $N_a$ we must calculate Christoffel symbols using 
the non-zero components of $q_{ab}$ 
expressed in Eq. ~(\ref{appmetric}).
The non-zero Christoffel symbols needed in the calculations are
\begin{mathletters}
\begin{eqnarray}
\Gamma ^1 _{11} = \Gamma ^2 _{12} &=& \Gamma ^3 _{13} = -\frac{Mx}{r^3} 
\left( 1+\frac{2M}{r} \right)^{-1} \ , \\
\Gamma ^2 _{11} = \Gamma ^2 _{33} &=& \frac{My}{r^3} \left( 1+\frac{2M}{r} 
\right)^{-1} \ , \\
\Gamma ^1 _{22} = \Gamma ^1 _{33} &=& \frac{Mx}{r^3} \left( 1+\frac{2M}{r} 
\right)^{-1} \ , \\
\Gamma ^2 _{22} = \Gamma ^1 _{12} &=& \Gamma ^3 _{23} = -\frac{My}{r^3} 
\left( 1+\frac{2M}{r} \right)^{-1} \ , \\
\Gamma ^1 _{13} = \Gamma ^2 _{23} &=& -\frac{Mz}{r^3} \left( 1+\frac{2M}{r} 
\right)^{-1} \ .
\end{eqnarray}
\end{mathletters}

By using the formula 
\begin{equation}
K_{ab} = \frac{1}{2N} \left( -\dot{q} _{ab} + N_{a|b} + N_{b|a} \right)
\end{equation}
for the exterior curvature tensor $K_{ab}$, we get
\begin{mathletters}
\begin{eqnarray}
K_{11}&=&\frac{2Jxy}{Nr^5}\left(\frac{2M}{r+2M}-3\right) \ , \\
K_{22}&=&-\frac{2Jxy}{Nr^5}\left(\frac{2M}{r+2M}-3\right) \ , \\
K_{33}&=&0 \ , \\
K_{12}=K_{21}&=&-\frac{J}{Nr^5}\left(x^2-y^2\right)\left(\frac{2M}{r+2M}-
3\right) \ , \\
K_{13}=K_{31}&=&\frac{Jyz}{Nr^5}\left(\frac{2M}{r+2M}-3\right) \ , 
\\
K_{23}=K_{32}&=&-\frac{Jxz}{Nr^5}\left(\frac{2M}{r+2M}-3\right) \ . 
\end{eqnarray}
\end{mathletters}
To evaluate $p_{ab}$, the canonical momentum conjugate to $q_{ab}$, from the 
expression
\begin{equation}
p_{ab}:=\frac{\sqrt{q}}{16\pi}\left(K_{ab}-q_{ab}K\right) \ ,
\label{appp}
\end{equation}
we must first calculate $K$ and $\sqrt{q}$, and we get
\begin{equation}
K:=q^{11}K_{11} + q^{22}K_{22} +q^{33}K_{33} = 0 \ , 
\label{appcurvature}
\end{equation}
\begin{equation}
\sqrt{q} = \left(1+\frac{2M}{r}\right)^{\frac{3}{2}} \ .
\label{appsqrtdet}
\end{equation}

When the results from Eqs.~(\ref{appcurvature}) and (\ref{appsqrtdet}) are substituted 
to Eq.~(\ref{appp}), we have
\begin{mathletters}
\begin{eqnarray}
p_{11}&=&\frac{\left(1+\frac{2M}{r}\right)^{\frac{3}{2}}}{16\pi}\frac{2Jxy}{Nr^
5}\left(\frac{2M}{r+2M}-3\right) \ , \\
p_{22}&=&-
\frac{\left(1+\frac{2M}{r}\right)^{\frac{3}{2}}}{16\pi}\frac{2Jxy}{Nr^5}\left(\frac{
2M}{r+2M}-3\right) \ ,  \\
p_{33}&=&0 \ ,  \\
p_{12}=p_{21}&=&-
\frac{\left(1+\frac{2M}{r}\right)^{\frac{3}{2}}}{16\pi}\frac{J}{Nr^5}\left(x^2-
y^2\right)\left(\frac{2M}{r+2M}-3\right) \ ,  \\
p_{13}=p_{31}&=&\frac{\left(1+\frac{2M}{r}\right)^{\frac{3}{2}}}{16\pi}\frac{Jy
z}{Nr^5}\left(\frac{2M}{r+2M}-3\right) \ ,  \\
p_{23}=p_{32}&=&-
\frac{\left(1+\frac{2M}{r}\right)^{\frac{3}{2}}}{16\pi}\frac{Jxz}{Nr^5}\left(\frac{2
M}{r+2M}-3\right) \ . 
\end{eqnarray}
\end{mathletters}
The boundary term $S^{\mathrm rev}_{\partial \Sigma}$ can be expressed by using the 
components calculated above and the components of 
the unit normal on the surface:
\begin{eqnarray}
S^{\mathrm rev}_{\partial \Sigma}&=&-2\int dt  \; \omega 
\oint\biggl[xq^{22}\left(p_{21}n^1+p_{22}n^2+p_{23}n^3\right)dS - 
 yq^{11}\left(p_{11}n^1+p_{12}n^2+p_{13}n^3\right)\biggl]dS \nonumber \\ 
 &=&\frac{1}{8 \pi}\int dt  \; \omega J 
\oint\left(1+\frac{2M}{r}\right)^{\frac{1}{2}}\left(\frac{2M}{r+2M}-
3\right)\frac{1}{Nr^5}
    \biggl[\left(x^2-y^2\right)\left(xn^1-yn^2\right) + \nonumber  \\
    &\ & 2xy\left(yn^1-xn^2\right) + z(x^2+y^2)n^3\biggl]dS \ .
\end{eqnarray}

This integral is easy to evaluate in spherical coordinates. We first consider a 2-dimensional spherical surface with radius
$r$. The relations between the spherical coordinates $r$, $\theta$ an $\phi$ and the 
Cartesian coordinates $x$, $y$ and $z$ are
\begin{equation}
x = r \cos \phi \sin \theta ,\ y = r \sin \theta \sin \phi , \ z = r \cos \theta
\end{equation}
The components of the unit normal $n^a, (a=1,2,3)$ on the surface are
\begin{equation}
n^1=n_x=\cos \phi \sin \theta ,\ n^2=n_y =\sin \theta \sin \phi , \ n^3=n_z = \cos \theta
\end{equation}
and the area element is
\begin{equation}
dS=r^2\sin\theta d\theta d\phi \ .
\end{equation}

In these coordinates the boundary term takes the form
\begin{eqnarray}
S^{\mathrm rev}_{\partial \Sigma} &=& \frac{1}{8 \pi}\int dt \; \omega J 
\left(1+\frac{2M}{r}\right)^{\frac{1}{2}}\left(\frac{2M}{r+2M}-3\right) 
    \int\limits^\pi _{\theta=0}\int\limits^{2\pi} _{\phi=0}
    \frac{1}{N}\biggl[\left(\cos^2 \phi - \sin^2 \phi \right)^2 \sin^5 \theta + \nonumber \\ 
  &\ & 4\cos^2 \phi \sin^2 \phi \sin^5 \theta + \cos^2 \theta \sin^3 \theta \biggl] d\phi 
d\theta
\end{eqnarray}

As $r$ approaches infinity $N$ goes to 1 and so the denominator can be approximated 
as 1.
Integration gives then:
\begin{equation}
S^{\mathrm rev}_{\partial \Sigma} \approx \frac{1}{8 
\pi}\left(1+\frac{2M}{r}\right)^{\frac{1}{2}}\left(
\frac{2M}{r+2M}-3\right)\frac{8\pi}{3}\int dt  \; \omega J  \ .
\end{equation}
and so the boundary term at infinity, where $r \rightarrow \infty$, is
\begin{equation}
S^{\mathrm rev}_{\partial \Sigma} = -\int dt \; \omega J \ .
\end{equation}

\section{A Novikov-type slicing of Kerr-Newman spacetime}
\label{app:B}

In this appendix we construct in details a slicing of Kerr-Newman spacetime in which the time coordinate of a freely falling observer through the bifurcation surface and the flat Minkowski time of a faraway observer at rest at the right hand side asymptotic infinity are identified. In a certain sense, one may view these observers and their time coordinates as physically equivalent. In Refs.~\cite{Louko} and~\cite{Mak2} similar identifications are performed and they are based on the Novikov coordinate system (see, for instance, Ref.~\cite{MTW}), where the time coordinate of a given point is given by the proper time $\tau$ of a freely falling observer in the Schwarzchild or Reisner-Nordstr\"om spacetime through that point, and the radial coordinate $R^\ast$ in the Novikov coordinate system is related to the point $r$ where the freely falling observer has begun his journey.

Since the $R$-coordinate in the classical Hamiltonian~(\ref{HinR}) can be geometrically interpreted as the radius of a wormhole throat at the equatorial plane $\theta =\pi /2$ in the Kerr-Newman black hole, we begin the construction of the slicing  with desired properties by considering the Kerr-Newman line element~(\ref{KNelement}) written in Boyer-Lindquist coordinates at the equatorial plane:
\begin{equation}
ds^2=-\frac{\Delta - a^2}{r^2}dt^2-\frac{2a(r^2+a^2-\Delta)}{r^2}dtd\phi+\frac{(r^2+a^2)^2-\Delta a^2}{r^2}d\phi^2
+\frac{r^2}{\Delta}dr^2 \ ,
\label{KNateq}
\end{equation}
where
\begin{equation}
\Delta=r^2+a^2-2Mr+Q^2\ .
\label{deltaag}
\end{equation}  
When the Boyer-Lindquist coordinates $x^\mu\ (\mu =0,1,2,3)$ satisfy the constraint
\begin{equation}
\sqrt{-g_{\mu \nu}{\dot x}^\mu{\dot x}^\nu}=1\ ,
\label{cons}
\end{equation}
where $g_{\mu\nu}$ gives the components of the metric tensor of the Kerr-Newman spacetime, the Lagrangian of a particle in the Kerr-Newman spacetime is, in general,
\begin{equation}
L_{\mathrm KN} =-\frac{1}{2}g_{\mu \nu}{\dot x}^\mu{\dot x}^\nu \ ,
\label{KNLag}
\end{equation}
At the equatorial plane Eqs.~(\ref{KNateq}) and~(\ref{KNLag}) give for the Lagrangian an expression 
\begin{equation}
L_{\mathrm KN} = \frac{1}{2}\frac{\Delta - a^2}{r^2}{\dot t}^2+\frac{a(r^2+a^2-\Delta)}{r^2}\dot t\dot \phi-\frac{1}{2}\frac{(r^2+a^2)^2
-\frac{1}{2}\Delta a^2}{r^2}{\dot \phi}^2-\frac{1}{2}\frac{r^2}{\Delta}{\dot r}^2\ ,
\end{equation}
where the dot denotes proper time derivative. The canonical momenta conjugate to $t$, $r$ and $\phi$ are:
\begin{mathletters}
\label{KNmomenta}
\begin{eqnarray}
p_t &=& \frac{\partial L_{\mathrm KN}}{\partial \dot t}=\frac{\Delta -a^2}{r^2}\dot t + \frac{a(r^2+a^2-\Delta)}{r^2}\dot \phi \label{KNmomentaa}\ ,\\
p_r &=& \frac{\partial L_{\mathrm KN}}{\partial \dot r}=-\frac{r^2}{\Delta}\dot r \label{KNmomentab}\ ,\\
p_\phi &=& \frac{\partial L_{\mathrm KN}}{\partial \dot \phi}=-\frac{(r^2+ a^2)^2-\Delta a^2}{r^2}\dot \phi + \frac{a(r^2+a^2-\Delta)}{r^2}\dot t \ .
\label{KNmomentac}
\end{eqnarray}
\end{mathletters}

It is easy to see that
\begin{mathletters}
\label{zero}
\begin{eqnarray}
\dot p_t &=& \frac{\partial L_{\mathrm KN}}{\partial  t}=0\ , \label{zeroa}\\
\dot p_\phi &\equiv& \frac{\partial L_{\mathrm KN}}{\partial  \phi}=0 \ .
\label{zerob}
\end{eqnarray}
\end{mathletters}
Therefore the momenta $p_t$ and $p_\phi$ are constants of motion. Let us denote these constants as
\begin{mathletters}
\label{constants}
\begin{eqnarray}
p_t&=:& \frac{R^*}{\sqrt{1+{R^*}^2}}\ , \label{constantsa}\\
p_\phi&=:& \ell\ ,
\label{constantsb}
\end{eqnarray}
\end{mathletters}
where we have introduced new real valued parameters $R^\ast$ and $\ell$.

The Hamiltonian  
\begin{equation}
H_{\mathrm KN}=p_r \dot r + p_t \dot t + p_\phi \dot \phi - L_{\mathrm KN}
\label{KNH}
\end{equation}
of a particle in Kerr-Newman spacetime can be shown to coincide with the Lagrangian $L_{\mathrm KN}$:
\begin{equation}
H_{\mathrm KN} = L_{\mathrm KN} \  .
\label{KNHL}
\end{equation}
Therefore, when the constraint~(\ref{cons}) is satisfied, we get
\begin{equation}
H_{\mathrm KN}= \frac{1}{2}\ . 
\label{KNHpuoli}
\end{equation}

On the other hand, when we use the parameters introduced in Eqs.~(\ref{constants}), we get from Eq.~(\ref{KNH}):
\begin{equation}
2H_{\mathrm KN}=1=-\frac{r^2}{\Delta}{\dot r}^2 +  \frac{R^*}{\sqrt{1+{R^*}^2}} \dot t + \ell \dot \phi\ ,
\label{2KNH}
\end{equation}
and from Eqs.~(\ref{KNmomentab}) and~(\ref{KNmomentac}) we obtain $\dot \phi$ and $\dot t$ in terms of $\ell$ and $R^\ast$: 
\begin{mathletters}
\begin{eqnarray}
\dot t &=& \frac{1}{\Delta}\left(\frac{r^4+2r^2a^2-a^2(\Delta-a^2)}{r^2}\frac{R^*}{\sqrt{1+{R^*}^2}}+\frac{a(r^2+a^2-\Delta)}{r^2}\ell\right) \  , \label{KNdott} \\
\dot \phi &=& \frac{1}{\Delta}\left(\frac{a(r^2+a^2-\Delta)}{r^2}\frac{R^*}{\sqrt{1+{R^*}^2}}-\frac{\Delta-a^2}{r^2}\ell\right)\  . \label{KNdotphi}
\end{eqnarray}
\end{mathletters}

Now, we choose $\ell = 0$, and because of this particular choice we get from Eq.~(\ref{2KNH}): 
\begin{equation}
2H_{\mathrm KN}=1=-\frac{r^2}{\Delta}{\dot r}^2+\frac{R^*}{\sqrt{1+{R^*}^2}} \frac{1}{\Delta}\left(\frac{r^4+2r^2a^2-a^2(\Delta-a^2)}{r^2}\frac{R^*}{\sqrt{1+{R^*}^2}}\right) \ .
\label{2KNH'}
\end{equation}
As we set $\dot r =0$, Eq.~(\ref{2KNH'}) yields us a quartic equation for $r$:
\begin{equation}
\Delta r^2=\frac{{R^*}^2}{1+{R^*}^2}\left[{r^4+2r^2a^2-a^2(\Delta-a^2)}\right] \ .
\label{rmax}
\end{equation}
From this equation one can calculate the $r$-coordinate $r_{\mathrm max}$ of the point from which an observer in a free fall begins his journey, in terms of $R^*$ which will henceforth be used as a radial coordinate of Kerr-Newman spacetime. Eq.~(\ref{2KNH'}) implies an implicit expression $r(\tau , R^*)$ for the "old" radial coordinate $r$ in terms of the "new" time coordinate $\tau$ and the "new" radial coordinate $R^*$:
\begin{equation}
\tau = \pm \sqrt{1+{R^*}^2} \int_{r_{\mathrm max}(R^*)}^{r(\tau, R^*)}\frac{r'^2}{\sqrt{{R^*}^2(r'^2+a^2)(2Mr'-Q^2)-r'^2\Delta}}\;dr' \ .
\label{inttau}
\end{equation}
In this equation the signs $+$ and $-$, respectively, correspond to the past and the future of the line where the time coordinate $t=0$ in the conformal diagram. To obtain an explicit expression $r(\tau, R^*)$ for $r$ one should first solve the quartic equation~(\ref{rmax}), and then perform the integration in Eq.~(\ref{inttau}). Solving Eq.~(\ref{rmax}), however, would yield a tremendously complicated expression for $r_{\mathrm max}$, and we shall not write it down here. However, it is easy to see that there are always at least two positive roots $r=r_{\mathrm max}=r_{\mathrm max}(R^*)$. This can be seen by plotting the both sides of Eq.~(\ref{rmax}) and varying $R^*$. Moreover, one finds that if one puts $r=r_{\mathrm max}=r_+$ then Eq.~(\ref{rmax}) implies $R^*=0$, and vice versa: if one sets $R^*=0$, then Eq.~(\ref{rmax}) is solved by $r=r_+$. Hence, we have found that for every $R^* \geq 0$ there is an observer in a free fall such that this observer is at rest at the time $t=\tau=0$ with respect to the "old" radial coordinate $r$. When $R^*=0$ our observer begins his journey at the bifurcation surface and his world line is a straight vertical line in the conformal diagram.

Can we extend this coordinate transformation to the right hand side asymptotic infinity? Yes we can, since we may choose the coordinate $R^\ast$ such that the solution $r= r_{\mathrm max}> r_+$ is the largest of the roots of Eq.~(\ref{rmax}). When this choice is made, one can show starting from Eq.~(\ref{rmax}), that for large $r_{\mathrm max}$
\begin{equation}
{R^*}^2 \sim -\left( 1+\frac{r_{\mathrm max}^2+a^2}{Q^2-2Mr_{\mathrm max}}\right)+{\mathcal O}(r_{\mathrm max}^{-3}) \ .
\label{large}
\end{equation}
Hence $R^*$ goes to infinity as $r_{\mathrm max}$ goes to infinity and vice versa. Moreover, the time coordinate $\tau$ of an observer at the asymptotic infinity coincides with the proper time $\tau$ of a freely falling observer at the wormhole throat.

Another matter to investigate still is that do the observers rotate or not with respect to the Boyer-Lindquist coordinates? We wrote our Hamiltonian from the point of view of an asymptotic non-rotating observer, and we assumed a foliation in which the time coordinate at the throat is a proper time of a non-rotating observer in a free fall. To show that in our foliation both of the observers are non-rotating we must show that $\dot \phi \rightarrow 0$ as $r\rightarrow r_+$ and $r\rightarrow \infty$. The latter case is straightforward, since in the expression 
\begin{equation}
\dot \phi = \frac{1}{\Delta}\frac{a(r^2+a^2-\Delta)}{r^2}\frac{R^*}{\sqrt{1+{R^\ast}^2}}\ ,
 \label{KNdotphi'}
\end{equation}
given by Eq.~(\ref{KNdotphi}) when $\ell =0$, the factor $\frac{R^*}{\sqrt{1+{R^*}^2}}$ approaches to one and the factor in front of it approaches to zero. The first case where $r\rightarrow r_+$ is a bit tricky, since we do not know the explicit relation of $r$ and $R^\ast$ at the bifurcation point. We have solved the tricky part by expanding Eq.~(\ref{rmax}) in terms of $r$ near the bifurcation point. If we take only the zeroth and the first order terms, we find that the point $r= r_{\mathrm max}$ where $\dot r=0$ is related to $R^\ast$ by an expression
\begin{equation}
r_{max}\approx \frac{r_+^5\left(2-\frac{{R^*}^2}{1+{R^*}^2}\right)-2Mr_+^4+a^2r_+^3\frac{{R^*}^2}{1+{R^*}^2}+4Ma^2r_+^2\frac{{R^*}^2}{1+{R^*}^2}-3a^2q^2r_+\frac{{R^*}^2}{1+{R^*}^2}}{2r_+^4\left(1-\frac{{R^*}^2}{1+{R^*}^2}\right)-2Mr_+^3+2Ma^2r_+\frac{{R^*}^2}{1+{R^*}^2} -2a^2q^2\frac{{R^*}^2}{1+{R^*}^2}} \ , 
\label{rmaxR*}
\end{equation} 
which  gives that $r_{max}=r_+$ as $R^*\longrightarrow 0$, as it should. Now, when Eq.~(\ref{rmaxR*}) is substituted into Eq.~(\ref{KNdotphi'}) and $R^\ast \rightarrow 0$, one gets the result
\begin{equation}
\dot \phi \longrightarrow 0\ .
\end{equation}
In other words, we have managed to construct a foliation of Kerr-Newman spacetime with desired properties at the equatorial plane: At the asymptotic infinity the time coordinate is the proper time of a freely falling, non-rotating observer at rest, and at the wormhole throat that of a similar non-rotating observer in a radial free fall through the bifurcation surface.

It is even possible to show that our construction gives the Novikov coordinate system in the Schwarzschild spacetime when one sets $q=a=0$ in Eq.~(\ref{rmax}). This result is given by Eqs.~(\ref{rmax}) and~(\ref{inttau}). We get an analogous coordinate system for the Reissner-Nordstr\"om spacetime when only $a=0$. It can be shown that then the relation between $r_{\mathrm max}$ and $R^\ast$ is 
\begin{equation}
r=r_{max}=\left(M+\sqrt{M^2-q^2{(1+{R^*}^2)}^{-1}}\right)(1+{R^*}^2)\ .
\end{equation}

\end{document}